\documentclass[11pt,a4paper]{article}
\pdfoutput=1
\usepackage{amsmath}
\usepackage{amsfonts}
\usepackage{amssymb}
\usepackage{graphicx}
\usepackage{mathrsfs}
\usepackage{amsfonts}
\usepackage{slashed}
\usepackage{pdfpages}
\usepackage{braket}
\usepackage{sidecap}
\usepackage[font=small,labelsep=none]{caption}
\usepackage{mathtools}
\usepackage{hyperref}
\newcommand{\be}{\begin{equation}}      
\newcommand{\ee}{\end{equation}}

\hypersetup{
    colorlinks=true,
    linkcolor=black,
    filecolor=black,      
    urlcolor=black,
    citecolor=black,
}
 
\newcommand{\R}{R}
\def\MM{M_{*}}
\def\MP{M_{\rm Pl}}

\newcommand*{\Ccc}{\mathscr }%
\newcommand{\vect}[1]{\boldsymbol{#1}}

\newcommand{\eqn}[1]{eq.~(\ref{#1})}
\newcommand{\secref}[1]{Sec.~\ref{#1}}

\usepackage[left=2cm,right=2cm,top=2.5cm,bottom=2.5cm]{geometry}
\author{Giovanni}
\begin{document}

\begin{center}
\LARGE{\bf Gravitational Wave Decay into Dark Energy}
\\[1cm] 

\large{Paolo Creminelli$^{\rm a}$,  Matthew Lewandowski$^{\rm b}$, Giovanni Tambalo$^{{\rm c},{\rm d}}$, Filippo Vernizzi$^{\rm b}$}
\\[0.5cm]

\small{
\textit{$^{\rm a}$
ICTP, International Centre for Theoretical Physics\\ Strada Costiera 11, 34151, Trieste, Italy}}
\vspace{.2cm}

\small{
\textit{$^{\rm b}$ Institut de physique th\' eorique, Universit\'e  Paris Saclay, CEA, CNRS \\ [0.05cm]
 91191 Gif-sur-Yvette, France}}
\vspace{.2cm}

\small{
\textit{$^{\rm c}$ SISSA, via Bonomea 265, 34136, Trieste, Italy}}
\vspace{.2cm}

\small{
\textit{$^{\rm d}$ INFN, National Institute for Nuclear Physics \\  Via Valerio 2, 34127 Trieste, Italy}}
\vspace{.2cm}

\end{center}

\vspace{0.3cm} 

\begin{abstract}\normalsize
We study the decay of gravitational waves into dark energy fluctuations $\pi$, through the processes $\gamma \to \pi\pi$ and $\gamma \to \gamma \pi$, made possible by the spontaneous breaking of Lorentz invariance. Within the EFT of Dark Energy (or Horndeski/beyond Horndeski theories) the first process is large for the operator $\frac12 \tilde m_4^2(t) \, \delta g^{00}\, \left(  {}^{(3)}\!\R  +  \delta K_\mu^\nu \delta K^\mu_\nu -\delta K^2  \right)$, so that the recent observations force $\tilde m_4 =0$ (or equivalently $\alpha_{\rm H}=0$). This constraint, together with the requirement that gravitational waves travel at the speed of light, rules out all quartic and quintic GLPV theories.  Additionally, we study how the same couplings affect the propagation of gravitons at loop order. The operator proportional to $\tilde m_4^2$ generates a calculable, non-Lorentz invariant higher-derivative correction to the graviton propagation. The modification of the dispersion relation provides a  bound on $\tilde m_4^2$ comparable to the one of the decay. Conversely, operators up to cubic Horndeski do not generate sizeable higher-derivative corrections.    

\end{abstract}

\vspace{0.3cm} 

\tableofcontents
\vspace{0.3cm}

\section{Introduction}

The observation of the gravitational wave event GW170817 \cite{TheLIGOScientific:2017qsa} and its electromagnetic counterpart \cite{Goldstein:2017mmi} started the detailed study of the propagation of gravitational waves (GWs). At variance with a cosmological constant---the simplest explanation of the present acceleration---models of Dark Energy (and Modified Gravity) act as a sort of ``medium", through which GWs travel. In the same way one uses the propagation of electromagnetic waves to study a material, 
GWs propagating through Dark Energy (DE) can be used to test these theories. Like a normal material, DE defines a preferred frame and thus spontaneously breaks Lorentz invariance. This implies that in general the speed of GWs may be different from the speed of light \cite{Lombriser:2015sxa,Bettoni:2016mij}. The recent observations put severe bounds on this possibility and therefore strong constraints on some DE models \cite{Creminelli:2017sry,Sakstein:2017xjx,Ezquiaga:2017ekz,Baker:2017hug}. See also \cite{Ezquiaga:2018btd}---and references therein---for a review.

In this paper we want to study another phenomenon that is possible due to the breaking of Lorentz invariance: the decay of gravitational waves into DE fluctuations. In a Lorentz invariant theory, a massless particle can only decay into two or more massless particles with all momenta exactly aligned. Measurable quantities must be summed over these collinear emissions to get rid of spurious IR divergences \cite{Lee:1964is,Weinberg:1995mt}. Once Lorentz invariance is broken, the excitations of DE will in general move at a speed different from the one of gravitons and the decay is allowed. For other works studying the damping of gravitational waves see e.g.~\cite{Deffayet:2007kf,Calabrese:2016bnu,Visinelli:2017bny,Amendola:2017ovw,Belgacem:2017ihm,Pardo:2018ipy}. 

We will study the decay of gravitons in the framework of the Effective Field Theory (EFT) of Dark Energy \cite{Creminelli:2006xe,Cheung:2007st,Creminelli:2008wc,Gubitosi:2012hu,Gleyzes:2013ooa}. We review the formalism in Sec.~\ref{sec:EFTDE} and specify it to the subset of theories with GWs travelling at the speed of light, since the others are not compatible with the recent data (assuming that the regime of validity of the EFT of DE encompasses the LIGO/Virgo scales \cite{deRham:2018red}). The connections with the covariant formalism, i.e.~to Horndeski and beyond Horndeski theories is treated in App.~\ref{App:Horndeski}. In App.~\ref{App:Frame} we discuss the invariance of the results under a disformal transformation. 

In Sec.~\ref{sec:decayrate} we derive  the cubic coupling $\gamma\pi\pi$ (where $\pi$ describes the DE fluctuations) and compute the decay rate of the process $\gamma \to \pi \pi$. It turns out to be very large  and thus incompatible with observations, for a particular operator of the EFT of DE: $\frac12 \tilde m_4^2(t) \, \delta g^{00}\, \left(  {}^{(3)}\!\R  +  \delta K_\mu^\nu \delta K^\mu_\nu -\delta K^2  \right)$. This conclusion holds if $\tilde m_4$ is large enough to play any role in modifying gravity and potentially affecting large-scale structure measurements. In the framework of Gleyzes-Langlois-Piazza-Vernizzi (GLPV) theories \cite{Gleyzes:2014dya,Gleyzes:2014qga}, setting $\tilde m_4 = 0$ (or equivalently $\alpha_{\rm H}=0$, where $\alpha_{\rm H}$ is defined in eq.~\eqref{alphaH}) and requiring GWs to travel at the speed of light corresponds in the covariant language to restricting to Horndeski up to the cubic Lagrangian (in particular no beyond Horndeski terms survive).
In the main text the calculations are done in Newtonian gauge, while in App.~\ref{app:flatgauge} they are done in the spatially flat gauge. We relegate to App.~\ref{app:gamma2pi} the derivation of the coupling $\gamma\gamma\pi$ and the computation of the decay rate of $\gamma \to \gamma \pi$, since this turns out to be subdominant.

The coupling $\gamma\pi\pi$ can be used to make a loop with external $\gamma$ legs: in other words, as we study in Section \ref{sec:loops}, at one-loop the graviton propagator is corrected. The calculable, i.e.~log-divergent, corrections give a sizeable dispersion of gravitational waves: a higher-dimension operator that is quadratic in the graviton and violates Lorentz-invariance is generated. Also this effect can be used to rule out the operator proportional to $\tilde m_4$. We study in general the radiative generation of higher dimension operators that can correct the graviton propagation. This allows one to rule out the operator $\tilde m_4$ even when the speed of $\pi$ is larger or equal to the speed of GWs and the decay $\gamma \to \pi \pi$ is impossible. The remaining theories, corresponding to Horndeski up to cubic order, do not generate sizeable higher derivative corrections. Conclusions and future directions are discussed in Section \ref{sec:conclusions}.

\section{\label{sec:EFTDE}EFT of Dark Energy
}

In this section we briefly introduce the EFT of DE and discuss the quadratic action of scalar and tensor perturbations.

\subsection{The EFT action}

To parametrize the interactions between gravitons and DE fluctuations we adopt the EFT description \cite{Creminelli:2006xe,Cheung:2007st,Creminelli:2008wc,Gubitosi:2012hu,Gleyzes:2013ooa}. 
This is particularly convenient to study fluctuations around cosmological FRW solutions with a preferred slicing induced by the time-dependent background scalar field.

In the usual unitary gauge, where  time coincides with uniform-field hypersurfaces, the EFT action expanded around a flat FRW background, $\text d s^2 = -\text d t^2 + a^2(t) \text d \vect x^2$, reads \cite{Creminelli:2017sry,Cusin:2017mzw}
\be
\begin{split}
\label{total_action}
S  =  & \int  \text d^4 x \sqrt{-g}   \bigg[  \frac{\MM^2}{2} f(t) \, {}^{(4)}\!R - \Lambda(t)- c(t) g^{00}  +  \frac{m_2^4(t)}{2} (\delta g^{00})^2 - \frac{m_3^3(t)}{2} \, \delta K \delta g^{00}  
- m_4^2(t) \delta {\cal K}_2 \\ 
& +\frac{\tilde m_4^2(t)}{2} \,\delta g^{00}\, {}^{(3)}\!\R  - \frac{m_5^2(t)}{2}  \delta g^{00} \delta {\cal K}_2  -   \frac{m_6(t)}{3} \delta {\cal K}_3  
 - \tilde m_6 (t) \delta g^{00} \delta {\cal G}_2  -   \frac{m_7(t)}{3} \delta g^{00} \delta {\cal K}_3  \bigg] \;,
\end{split}
\ee
with
\be
\label{deltaKK}
\begin{split}
\delta {\cal K}_2 &\equiv \delta K^2 -  \delta K_\mu^\nu \delta K^\mu_\nu \;, \qquad \delta {\cal G}_2  \equiv  \delta K_\mu^\nu \, {}^{(3)}\!R^\mu_\nu -  \delta K \, {}^{(3)}\!R/2 \;, \\
\delta {\cal K}_3  &\equiv \delta K^3 -3 \delta K \delta K_\mu^\nu \delta K^\mu_\nu + 2 \delta K_\mu^\nu \delta K^\mu_\rho \delta K_\nu^\rho   \;.
\end{split}
\ee
In the above action, ${}^{(4)}\!R$ is the 4d Ricci scalar, $\delta g^{00} $ denotes the perturbation of $g^{00}$ around the background solution, $\delta g^{00}  \equiv 1+g^{00}$, $\delta K_\mu^\nu\equiv K_\mu^\nu- H \delta_\mu^\nu$  is the perturbation of the extrinsic curvature of the equal-time hypersurfaces, with $H\equiv \dot a/a$ being the Hubble rate, $\delta K$ its trace, and  ${}^{(3)}\! R$ is the 3d Ricci scalar of these hypersurfaces. For notational convenience, we have defined the quantities in eq.~\eqref{deltaKK}, where  
${}^{(3)}\!R_\mu^\nu$ is the 3d Ricci tensor of  the time hypersurfaces.
While $\MM^2$ is a constant, the other parameters are slowly-varying time-dependent functions.
To have sizable effects for structure formation, one typically considers $m_2^4\sim \MP^2 H_0^2$, $m_3^3 \sim \MP^2 H_0$, $m_4^2 \sim \tilde m_4^2 \sim m_5^2 \sim \MP^2$ and $m_6 \sim \tilde m_6 \sim m_7 \sim \MP^2 H_0^{-1}$, where $\MP$ is the Planck mass. Notice also that positive powers of the mass-dimension quantities $m_i$ and $\tilde m_i$  can have either sign.

This action governs the  cosmological evolution in Horndeski \cite{Horndeski:1974wa,Deffayet:2011gz} (obtained for $\tilde m_4^2=m_4^2$ and $\tilde m_6=m_6$) and GLPV theories. The first three operators are sufficient to describe the background evolution, while the following four describe linear perturbations. The other terms describe non-linearities; for simplicity we have written only those  that contribute to the leading number of spatial derivatives: these operators dominate  the nonlinear regime of structure formation and  the Vainshtein regime (see e.g.~\cite{Kimura:2011dc,Kobayashi:2014ida,Cusin:2017wjg,Dima:2017pwp} for details). The relation between the EFT parameters and the Horndeski and beyond Horndeski functions is given in App.~\ref{App:Horndeski}.

To be compatible with the constraints  from the GW170817 event \cite{Monitor:2017mdv}, in the following we will assume that  gravitational waves propagate at the speed of light, $c_\text{T}=1$.
Since in a generic theory the speed of tensors computed around the cosmological background is \cite{Gleyzes:2013ooa}
\be
\label{cT}
c_\text{T}^2  = 1 - \frac{2 m_4^2}{\MM^2 f +2 m_4^2} \;, 
\ee 
in the ``frame'' where photons and gravitons propagate on the light-cone the action simplifies as one has  
\be
\label{m4zero}
m_4^2 = 0 \;. 
\ee
A further simplification comes from requiring that this speed is stable to small changes of the background. As discussed in \cite{Creminelli:2017sry}, this implies  
\be
\label{mpercT1}
\tilde m_4^2 = m_5^2  \;, \qquad m_6 = \tilde m_6 = m_7 =0 \;.
\ee
Finally, $f$ can be set to be constant by a conformal transformation of the metric, which do not change the speeds of propagation.
In general, the conformal transformation changes the couplings between matter and the DE field but the interactions between gravitons and DE do not depend on these matter couplings. Therefore, there is no loss of generality in choosing this frame. We will discuss the constraints in a more generic frame in App.~\ref{App:Frame}.

With the above assumptions and this last simplification, the unitary-gauge action becomes
\be
\begin{split}
\label{starting_action}
S  =  \int  \text d^4 x \sqrt{-g}  \bigg[ & \frac{\MP^2}{2}  \, {}^{(4)}\!R - \Lambda(t)- c(t) g^{00}  +  \frac{m_2^4(t)}{2} (\delta g^{00})^2  \\ 
&- \frac{m_3^3(t)}{2} \, \delta K \delta g^{00}  
 +\frac{\tilde m_4^2(t)}{2} \, \delta g^{00}\, \left(  {}^{(3)}\!\R   -   \delta {\cal K}_2   \right) \bigg] \;,
\end{split}
\ee
where the (time-independent) Planck mass squared is $\MP^2 = \MM^2 f$. For simplicity, in the following we will assume that the mass scales $m_3^3$ and $\tilde m_4^2$ are time independent but taking into account their slow time dependence is straightforward.

Before expanding the action we note that, on the homogeneous solution, the variation of eq.~\eqref{starting_action} with respect to the metric yields
two equations that can be used to express $c(t)$ and $\Lambda(t)$ in terms of the Hubble expansion and matter quantities (see e.g.~\cite{Cheung:2007st,Gubitosi:2012hu}). To remain general, we refrain from giving their precise expressions here because these equations depend on the details of the matter Lagrangian, e.g., the matter coupling with the scalar, and these are irrelevant for the following discussion. We will therefore treat $c(t)$ and $\Lambda(t)$ as independent functions.

As it will be made explicit below, the unitarity cutoff of the EFT of DE is usually of order
\be
\Lambda_3 \equiv (\MP H_0^2 )^{1/3} \;.
\ee
This scale corresponds to roughly $1000$ km and therefore is within the LIGO/Virgo frequency band. If new physics enters around the scale $\Lambda_3$, one expects at most  ${\cal O}(1)$ corrections to the predictions of the EFT. On the others hand, if new states are present at a scale parametrically below $\Lambda_3$, the EFT is of no use for the gravitational waves predictions at LIGO/Virgo \cite{deRham:2018red}. In the following we will assume the validity of the EFT description.

\subsection{Free theory}

We will first   expand the  action at quadratic order  and then derive the graviton-scalar interactions in Sec.~\ref{sec:decayrate}.
For later convenience, 
it is useful to use the  standard ADM metric decomposition, where the metric line element reads 
\be
\text ds^2 = - N^2 \text dt^2 + h_{ij} (\text dx^i + N^i \text dt) (\text dx^j + N^j \text dt) \;. 
\ee
In terms of this decomposition, the extrinsic curvature takes the form
\begin{equation}
K_{ij} = \frac{1}{N} E_{ij} \;, \qquad E_{ij} \equiv \frac12 (  \dot{h}_{ij} - D_i N_j - D_j N_i ) \;, 
\end{equation}
where $D_i$ is the covariant derivative with respect to the metric $h_{ij}$, which is also used to raise and lower Latin indexes. 
In this section we will work in  Newtonian gauge, defined by
\be
\label{eq:gaugeNewton}
N^2 = 1+2 \Phi \;, \qquad N_i = 0 \;, \qquad h_{ij} = a^2(t) (1-2 \Psi) (e^\gamma)_{ij} \;,
\ee
with  $\partial_i \gamma_{ij} = \gamma_{ii} = 0$. The derivation of the quadratic action and the graviton-scalar interactions in spatially flat gauge is left to App.~\ref{app:flatgauge}.

The time-diffeomorphism invariance of the action can be restored by the usual Stueckelberg trick (see e.g.~\cite{Cheung:2007st,Gleyzes:2013ooa}). By defining the Goldstone bosons of  broken time diffeomorphisms by $\pi$, under a time coordinate change $t \to t+ \pi(t,\vect x)$ we obtain
\begin{align}
g^{00} & \to g^{00} + 2 g^{0 \mu} \partial_\mu \pi + g^{\mu \nu} \partial_\mu \pi \partial_\nu \pi \;, \label{g00stueck}\\
\delta K_{i}^j & \to \delta K_{i}^j - \dot H \pi \delta_i^j - \frac{1}{a^2} \partial_i \partial_j \pi + {\cal O}(2)\;,  \label{Kijstueck}\\
{}^{(3)}\!\R & \to {}^{(3)}\!\R+ \frac{4}{a^2} H \partial^2 \pi + {\cal O}(2) \;,\label{3Rstueck}
\end{align}
where, since $\delta K_{i}^j$ and ${}^{(3)}\!\R$ enter in the action only multiplied by a perturbation, in the last two equations we have kept only linear perturbations and used that $\partial^2 \equiv \sum_i\partial_i \partial_i$.

Varying the action with respect to $\Phi$ and focussing on the sub-Hubble limit by keeping only the leading terms in spatial derivatives, one obtains
\be\label{Phieq}
2 \MP^2 \partial^2 \Psi + m_3^3   \partial^2 \pi + 4 \tilde m_4^2 \partial^2 ( \Psi + H \pi)  = 0  \;,
\ee
which can be solved for $\Psi$ in terms of $\pi$,
\be
\label{phisol}
\Psi  = - \frac{m_3^3 + 4 \tilde{m}_4^2 H}{2(\MP^2 + 2 \tilde{m}_4^2 )}  \pi \;.
\ee
Variation with respect to $\Psi$ in the same limit yields
\be\label{Psieq}
 \MP^2 \partial^2 (\Phi - \Psi)  + 2 \tilde m_4^2 \partial^2 ( \Phi - \dot \pi)  = 0  \;.
\ee
Since the frequencies involved in the gravitational wave experiments that concern us here are much higher than the Hubble rate, one can focus on the highest number of time derivatives per field (i.e. we assume $H \pi \ll \dot \pi$) and express $\Phi$ in terms of $\dot \pi$,
\be
\label{Phitopidot}
\Phi  = \frac{2 \tilde{m}_4^2}{\MP^2 + 2 \tilde{m}_4^2} \dot \pi  \;.
\ee
Plugging these solutions back into the action one obtains (see also \cite{Gleyzes:2013ooa})
\begin{equation}\label{S2pi}
S_\pi^{(2)}  = \int \text d^4 x \MP^2\frac{ 3 m_3^6  + 4 \MP^2 (c + 2 m_2^4)}{4(\MP^2 + 2 \tilde{m}_4^2)^2}\left[\dot{\pi}^2 - {c_s^2}  (\partial_i \pi )^2  \right] \;,
\end{equation}
where $c_s^2$ is the speed of sound squared, which is given by
\begin{equation}
\label{csquared}
c_s ^2 =\frac{4 \left(\MP^2+2 \tilde{m}_4^2\right)^2 c-\MP^2 \left(m_3^3-2 \MP^2 H\right) \left(m_3^3+4 \tilde{m}_4^2 H\right)+ 8\MP^2 \tilde{m}_4^2 \left(\MP^2+2 \tilde{m}_4^2\right)\dot H  }{M_{\text{Pl}}^2\left[3 m_3^6 +4  \MP^2 (c + 2 m_2^4)\right]} \;.
\end{equation}
Once again, in this paper we  consider frequencies much higher than $H$: we can assume that we are in Minkowski spacetime and set  $a=1$, as we did for eq.~\eqref{S2pi}.

A comment is in order here. 
It is a known peculiar feature of beyond Horndeski theories that the dynamics of $\pi$ is affected by the mixing with matter fluctuations \cite{Gleyzes:2014dya,DAmico:2016ntq}. However, this mixing is neglected in eq.~\eqref{S2pi} by neglecting matter fluctuations in its derivation, in eqs.~\eqref{phisol} and \eqref{Phitopidot}.  
This is justified by the fact that the mixing would depend on the local environment and on scales of order $1000$~km one cannot rely on small perturbations around the cosmological average value. Since the mixing depends on the position, in the following the coefficients of the $\pi$ action, in particular the speed of sound, should be considered as weakly position dependent. This approximation does not change our conclusions, however. (Neglecting matter fluctuations becomes exact in the limit in which DE dominates in the Friedmann equations.) 

Getting back to eq.~\eqref{S2pi}, we define the canonically normalized field $\pi^{(c)}$ as 
\begin{equation}\label{eq:pican}
\pi^{(c)} \equiv  \frac{\MP \big[  3 m_3^6 + 4 \MP^2 (c + 2 m_2^4) \big]^{\frac{1}{2}}}{\sqrt{2}(\MP^2 + 2 \tilde m_4^2)}\pi \;.
\end{equation}

The quadratic action for the graviton can be  found by expanding the Einstein-Hilbert term. One gets
\begin{equation}
S_\gamma^{(2)} = \int \text d^4 x \frac{\MP^2}{8}\left[\dot{\gamma}_{ij}^2 -  ( \partial_k \gamma_{ij} )^2  \right] \;.
\end{equation}
Defining the Fourier decomposition of $\gamma_{ij}$ as 
\begin{equation}
\gamma_{ij}(t,\vect{x}) =  \int \frac{\text d ^3 \vect k}{(2 \pi)^3}  \sum_{\sigma = \pm} \epsilon_{ij}^\sigma(\vect{k}) \gamma_{\vect{k}}^{\sigma}(t)  e^{i \vect{k} \cdot \vect x} \;,
\end{equation}
where $+$ and $-$ are the two polarizations of the graviton, with
\begin{equation}
\epsilon_{ij}^\sigma(\vect{k}) \delta^{ij}  = k^i \epsilon_{ij}^\sigma (\vect{k}) = 0 \;, \qquad  \epsilon_{ij}^\sigma(\vect{k}) \epsilon_{ij}^{\star \sigma'}(\vect{k}) = 2 \delta_{\sigma \sigma'} \;,
\end{equation}
the canonical normalized Fourier modes of the graviton are  
\begin{equation}\label{eq:gammacan}
\gamma_{ij}^{(c)} \equiv \frac{\MP}{\sqrt{2}}\gamma_{ij} \;.
\end{equation}

For later convenience, we note that the tensor product of two polarizations has to be transverse in each of its indexes and traceless in two couples of indexes. It is thus given  by
\begin{align}
\label{prodeps}
\sum_{\sigma= \pm} \epsilon_{ij}^\sigma(\vect{k}) \epsilon^{\star \, \sigma}_{mn}(\vect{k}) =  \lambda_{im}\lambda_{jn}  + \lambda_{in} \lambda_{jm} - \lambda_{ij}\lambda_{mn}  \;, \qquad 
& \lambda_{ij} \equiv \delta_{ij} - \frac{k_i k_j}{\vect{k}^2} \;.
\end{align}

\section{Graviton decay into $\pi \pi$}
\label{sec:decayrate}
As mentioned in the introduction, the dominant decay channel is the decay of gravitational waves into two scalar fluctuations. In this section we compute the interaction vertex and the rate associated to this decay.

\subsection{Interaction vertex $\gamma \pi \pi$}

Let us compute the cubic vertex of the interaction $\gamma \pi \pi$ in the gauge specified in \eqref{eq:gaugeNewton}. We first inspect the Einstein-Hilbert term in the action \eqref{starting_action}, to see if it can generate such a coupling. 
Since the 4d Ricci scalar is 4d diffeomorphism invariant, we do not need to perform the Stueckelberg trick on it. We can decompose it in the $3+1$ quantities using the Gauss-Codazzi relation, i.e.
\begin{equation}\label{eq:EHaction}
S_{\text{EH}} = \frac{\MP^2}{2}\int \text d ^4 x \sqrt{h} \left[N\; {}^{(3)} \! R + N^{-1} \left(E_{ij} E^{ij}  - E^2 \right)\right] \;.
\end{equation}
One can verify that the 3d scalar quantities $\sqrt{h}$, $E$ and $E_{ij} E^{ij}$ do not yield any  contribution  linear  in $\gamma_{ij}$. While $ {}^{(3)} \! R$ gives a term linear in $\gamma_{ij}$, this contains fewer derivatives than the terms discussed below.
Therefore, we disregard $S_{\text{EH}}$.

Discarding the operators proportional to $\Lambda$, $c$, $m_2^4$ and $m_3^3$, which do not contain linear terms in $\gamma_{ij}$, we focus on the operator proportional to $\tilde m_4^2$, whose contribution to the action is
\begin{equation}\label{eq:term4}
 S_4 = \frac{\tilde{m}_4^2}{2}\int  \text d^4 x \, N \sqrt{h}\, \delta g^{00} \left[ {}^{(3)}\! R  +  \delta K_{ij} \delta K^{ij} - \delta K ^2\right] \;.
\end{equation}
For the 3d Ricci, eq.~\eqref{3Rstueck} is not enough, since  we need to perform the  Stueckelberg trick at linear order in both $\pi$ and in $\gamma$. 
Starting from the linear expression $^{(3)} \! R = \partial_i \partial_j h_{ij} -\partial^2 h$ and using the following transformations under a time-deffeomorphism,  
\begin{align}
h_{ij} & \to h_{ij} - N_i \partial_j \pi - N_j \partial_i \pi + {\cal O}(\pi^2) \;, \\
\partial_i & \to \partial_i - \partial_i\pi \partial_0 + {\cal O}(\pi^2) \;,
\end{align}
one gets, neglecting the expansion of the universe, 
\be\label{3Rstuck}
{}^{(3)}\! R \to  {}^{(3)}\! R - 2 \partial_i \partial_j (N_i \partial_j \pi) + 2 \partial^2 (N_i \partial_i \pi) - \partial_i \pi \partial_j \dot h_{ij} - \partial_i (\partial_j \pi \dot h_{ij} ) + \partial_i (\partial_i \pi \dot h) + \partial_i \pi \partial_i \dot h  \;,
\ee 
which in our gauge becomes 
\be
\label{3Rgen}
{}^{(3)}\! R \to  {}^{(3)}\! R - \dot\gamma_{ij} \partial_i \partial_j \pi  \;.
\ee
By multiplying by $\delta g^{00}$  after the Stueckelberg trick  (see eq.~\eqref{g00stueck}) this term generates the following contribution to the action
\be
\label{1cont}
 - \int \text d^4 x\,  \frac{\tilde{m}_4^2}{2} (2 \Phi - 2 \dot{\pi})\dot \gamma_{ij}\partial_i \partial_j\pi \;,
\ee
where we have retained only terms with the highest number of time derivatives.

For the terms quadratic in the extrinsic curvature in the bracket of eq.~\eqref{eq:term4}, it is enough to use the linear Stueckelberg trick, eq.~\eqref{Kijstueck}. While $\delta K^2$ does not generate  terms linear in $\gamma_{ij}$ unsuppressed by $H$, $\delta K_{ij} \delta K^{ij}$ generates $- \dot\gamma_{ij} \partial_i \partial_j \pi$.
Multiplying by $\delta g^{00}$, this gives an identical contribution as eq.~\eqref{1cont}.  
Replacing $\Phi$ using eq.~\eqref{Phitopidot} and integrating by parts, we finally obtain
\begin{align}
\label{gammapipi}
S_{\gamma \pi \pi} = \frac{\MP^2 \tilde m_4^2}{\MP^2 + 2 \tilde m_4^2} \int \text d^4 x\,  \ddot \gamma_{ij} \partial_i \pi \partial_j \pi \;.
\end{align}

Using the canonically normalized fields defined in eqs.~\eqref{eq:pican} and \eqref{eq:gammacan}, the interaction vertex \eqref{gammapipi}
becomes
\begin{equation}
L_{\gamma \pi \pi} =  \frac{1}{\Lambda_*^3}\ddot{\gamma}_{ij}^{(c)} \partial_i \pi^{(c)} \partial_j \pi^{(c)}  \;, 
\end{equation}
with 
\begin{equation}
\label{Lambdadef}
\Lambda_*^3 \equiv \MP \frac{3 m_3^6 + 4 \MP^2(c+2 m_2^4)}{2\sqrt{2}\, \tilde{m}_4^2 (\MP^2 + 2 \tilde{m}_4^2 )} \;.
\end{equation}
In the following we denote by $ p^\mu$, $ k_1^\mu$ and $ k_2^\mu$, respectively the 4-momentum of the decaying graviton and of the two $\pi$ fields in the final state. Therefore, in diagrammatic form in Fourier space, for a given polarization $\sigma$ the interaction vertex reads 
\begin{equation}\label{eqVertex2}
\vcenter{\hbox{\includegraphics[width=0.2\textwidth]{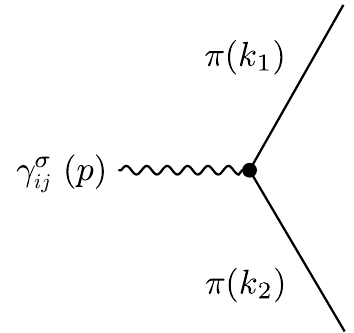}}}=  2 \times  \frac{1 }{\Lambda_*^3} p^2\, k_{1m}  k_{2n}\left[ \frac{1}{2}(\delta_{i m} \delta_{j n} + \delta_{i n} \delta_{j m}) - \frac{1}{3}\delta_{ij}\delta_{mn}\right] \;,
\end{equation}
where the factor of $2$ comes from the two possibilities of associating $k_1$ and $k_2$.

\subsection{Decay rate}
\label{subsec:decayrate}

Let us define the matrix element $i \Ccc A$  for a given polarization state $\sigma$ as
\begin{equation}
\label{defA}
  \braket{\{p, \sigma\}; \text{in} | k_1,k_2; \text{out}} \equiv (2 \pi)^4 \delta^{(4)}(p^\mu - k_1^\mu - k_2^\mu) \, i \Ccc A \;.
\end{equation}
The  decay rate  
reads
\be
\label{decayrate}
\Gamma_{ \gamma \rightarrow \pi \pi} =  \frac{1}{2} \times \frac{1}{2 E_p} \int \frac{\text d ^3 \vect k_1}{(2 \pi)^3 2 E_{k_1}}\frac{\text d ^3 \vect  k_2}{(2 \pi)^3 2 E_{k_2}} (2 \pi)^4 \delta^{(4)}(p^\mu - k_1^\mu - k_2^\mu) \braket{|i \Ccc A|^2} \;,
\ee
(the factor $1/2$ in front of the integral  comes from considering identical final particles) where, for any 4-vector $q^\mu$, $E_q$ denotes its time component
 and $\braket{|i \Ccc A|^2}$ is the square of the matrix element $i \Ccc A$  averaged over all possible initial polarizations for the in-state.
Before evaluating this explicitely, we can simplify the integral. 

Integrating over $\text d^3 \vect k_2$ removes   $\delta^{(3)} (\vect p - \vect  k_1 - \vect k_2)$. 
Then, let us define  $p \equiv | \vect{p}|$, $k_1 \equiv | \vect{k}_1|$ and $k_2 \equiv | \vect{k}_2|$.
Integrating over $\text d k_1$ 
using the on-shell conditions (we neglect the mass of $\pi$ assuming that it is much smaller than the typical frequency under consideration)
\be
E_p = p \;, \qquad E_{k_1} = c_s k_1 \;, \qquad E_{k_2} = c_s k_2 \;,
\ee
removes $\delta (E_p - E_{ k_1} - E_{ k_2})$. For this last step, it is convenient to define $\Omega \equiv {\vect{k}}_1 \cdot {\vect{p}}/(k_1 p)$ and express $k_2$ in terms of $k_1$ as 
\begin{equation}
\label{k2fork1}
k_2 = \sqrt{k_1^2 + p^2 - 2 p k_1 \Omega} \;.
\end{equation}
Finally, assuming $0 < c_s<1$ and expressing $k_1$ in terms of $p$ and $\Omega$ using 
\be
\label{k1inp}
k_1 = \frac{p(1-c_s^2)}{2c_s(1-c_s \Omega)} \;,
\ee
one obtains
\be\label{Gamma}
\Gamma_{ \gamma \rightarrow \pi \pi} =   \frac{1}{4 p}  \frac{1}{16 \pi c_s^3} \int_{-1}^{1} \text d \Omega \, \frac{1-c_s^2}{(1-c_s \Omega)^2} \braket{|i \Ccc A|^2} \;.
\ee

We can now compute $\braket{|i \Ccc A|^2}$. The matrix element of  the vertex \eqref{eqVertex2} is
\begin{equation}\label{matrixel}
i \Ccc A  = \frac{2 i}{\Lambda_*^3} p^2\,k_{1i}  k_{2 j}  \, \epsilon_{ij }^{\star \sigma}(\vect p) \;.
\end{equation}
Averaging over all possible initial polarizations for the in-state, using energy-momentum conservation, $k_2^\mu= p^\mu - k_1^\mu $, the transversality of the polarization tensor, $p^i  \epsilon_{ij }^{ \sigma}(\vect p) = 0$, and  eq.~\eqref{prodeps}, we find
\begin{align}
\label{Amplitude}
\braket{|i \Ccc A|^2} \equiv \frac{1}{2}\sum_{\sigma = \pm} |i \Ccc A|^2 = \frac{2 p^4}{\Lambda_*^6}  {k}_1^4\left(1- \Omega^2 \right)^2 \;.
\end{align}
Finally, replacing this expression  inside the integral and integrating over $\text d \Omega$ using eq.~\eqref{k1inp} we obtain
\begin{align}
\Gamma_{ \gamma \rightarrow \pi \pi} = \frac{p^7 (1-c_s^2)^2}{480 \pi c_s^7 \Lambda_*^6} \label{eq:decaygpp} \;.
\end{align}

For LIGO/Virgo observations we have $p \sim \Lambda_3$. Requiring that the 
gravitational waves are stable over cosmological distances $\sim H_0^{-1}$,
 one gets
\be
\label{limitdecay}
\frac{\Lambda_3}{H_0} \left( \frac{\Lambda_3}{\Lambda_*} \right)^6 \frac{(1-c_s^2)^2}{480 \pi c_s^7 } \sim 10^{20} \left( \frac{\Lambda_3}{\Lambda_*} \right)^6 \frac{(1-c_s^2)^2}{480 \pi c_s^7 } \lesssim 1\;,
\ee
which implies that $\Lambda_* \gg \Lambda_3$. 
To compare with large-scale structure constraints, we  can write the scale $\Lambda_*$ in terms of  quantities constrained by observations. In particular, let us define the dimensionless quantities \cite{Bellini:2014fua,Gleyzes:2014rba}
\be\label{alphaH}
\alpha \equiv \frac{2 c+ 4 m_2^4}{\MP^2 H^2} + \frac{3 m_3^6}{2 \MP^4 H^2}  
 \;, \qquad \alpha_{\rm B} \equiv  -\frac{m_3^3}{2 \MP^2 H}\;, \qquad  \alpha_{\rm H} \equiv \frac{2 \tilde m_4^2}{\MP^2} \;.
 \ee
The first quantity, $\alpha$, sets the normalization of the scalar fluctuations and must be positive while $\alpha_{\rm B}$ \cite{Bellini:2014fua} and $\alpha_{\rm H}$ \cite{Gleyzes:2014qga} measure the kinetic mixing of the scalar respectively with gravity and matter. One finds
\be
\left( \frac{\Lambda_3}{\Lambda_*} \right)^3= \frac{\alpha_{\rm H} (1+\alpha_{\rm H})}{\sqrt{2} \, \alpha}\;, 
\ee
so that, from eq.~\eqref{limitdecay}, $\alpha_{\rm H}$---and thus $\tilde m_4^2$---must vanish for any practical purpose. Notice that one cannot avoid this conclusion taking $\alpha$ very large: this limit corresponds to $c_s \ll 1$ and further enhances the decay rate eq.~\eqref{eq:decaygpp}. Moreover, in the same way one cannot take $\alpha_{\rm H}$ close to $-1$. Indeed, in this case the speed of sound squared becomes negative, as one can see from eq.~\eqref{csquared}, and the system is unstable.

For interesting values of $\tilde m_4^2$ the decay rate of the GWs is so large that no wave will reach the detector. For this reason it is not worthwhile to look at the precise effects on the luminosity distance as a function of the frequency. Concerning the produced scalar modes, these will not form a possibly detectable burst but they will be emitted in different directions and spread in space. Notice also that our perturbative calculation does not take into account the presence of a large number of quanta giving rise to a classical wave: coherent effects will further enhance the loss of energy into scalar waves.

Before concluding the section, let us briefly discuss the case $c_s \ge 1$. For $c_s^2 = 1$, energy-momentum conservation implies that the $\pi$'s are collinear with $\gamma$, i.e.~$\Omega =1$. In this configuration the decay is forbidden by the conservation of the angular momentum: the graviton with helicity 2 cannot decay collinearly into scalar particles. (Indeed, in this limit the interaction \eqref{matrixel} vanishes by the transversality of the graviton polarization.) Instead, the case $c_s>1$ is kinematically forbidden by energy-momentum conservation. We will discuss in the next section that also in the case $c_s \ge 1$ the operator proportional to $\tilde m_4^2$ must be negligibly small.

\section{\label{sec:loops}Loop corrections and dispersion}

We now move to study the loop corrections to the graviton propagator induced by the coupling $\gamma \pi \pi$. As argued in \cite{Creminelli:2017sry}, setting $c_\text T = 1$ is stable under quantum corrections. However, since Lorentz invariance is spontaneously broken, loop corrections could modify the dispersion relation of gravitons (i.e.~provide an energy-dependent phase-velocity) at a level in principle detectable by current gravitational waves experiments \cite{Yunes:2016jcc,LIGO2017}. The bounds on a possible non-trivial dispersion are even tighter than the ones on $c_{\rm T}$, since they rely on the comparison among different frequencies and are not limited by the astrophysical uncertainty on the emission time. Moreover, the result for the decay rate obtained in the previous section suggests that these dispersion effects are of conspicuous size. Indeed, it is well known that absorption is often accompanied by dispersion effects of the same magnitude. In this section we want to investigate these effects by computing loop-corrections to the graviton propagator and look at possible higher-derivative corrections.

\subsection{Graviton self-energy}
\label{sec:4.1}

As already done for the decay rate, we focus on the interaction vertex \eqref{eqVertex2}, which turns out to be the dominant coupling at the energy scales relevant for gravitational wave experiments. The corresponding term in the action can be cast in a manifestly 3-dimensional covariant form as $\sim \nabla^0 K_{\mu \nu} \partial^\mu \pi \partial^\nu \pi$. Therefore, operators generated at loop-level from this interaction do preserve diffeomorphism invariance.\footnote{Radiative corrections will generate terms that are manifestly invariant under time-dependent spatial diffs. In the vertex $\delta g^{00} \delta K_{ij} \delta K^{ij}$ one has an external $\delta K_{ij}$ leg, which is explicitly covariant under time-dependent spacial diffs. One can integrate by parts and move derivatives that act on the internal $\pi$'s on the external leg: this shows that the operator $\delta K_{ij}^2$ is not renormalized in compliance with the non-renormalization theorem of Galileons \cite{Luty:2003vm}. Since we are interested in the effect on the propagation of gravitational waves, we disregard spatial derivatives acting on $K_{ij}$: these will contribute to operators that depend on $\partial_i K_{ij}$, and these cannot affect gravitational waves, since they are transverse. The external leg can thus be taken of the form $\partial_0 K_{ij}$. (Invariance under time-dependent spatial diffs at all orders implies one gets a structure $\nabla^0 K_{ij}$; here we disregard higher order terms.)

Things are less transparent for the interaction $\delta g^{00} {^{(3)}} R$. In the calculation one has to take out of $^{(3)} R$ a gravitational wave and a scalar so that one cannot keep objects that are explicitly covariant under time-dependent spatial diffs. To check the invariance it is useful to look at the terms linear in $\pi$ that originate from $^{(3)} R$, eq.~\eqref{3Rstuck}. 
One can explicitly check the invariance of eq.~\eqref{3Rstuck} under time-dependent spatial diffs: $h_{ij} \to h_{ij} + \partial_i \xi_j + \partial_j \xi_i$, $N^i \to N^i + \partial_0 \xi^i$. In particular, since we are interested only in the effect on gravitational waves, one can disregard terms that vanish for transverse, traceless perturbations and focus on the two terms: $\partial_i \partial_j \pi (-\dot h_{ij} + 2 \partial_j N_i) = -\frac12 \partial_i \partial_j \pi K^{ij}$. The generated terms relevant for gravitational waves have the same structure as in the case $\delta g^{00} \delta K_{ij} \delta K^{ij}$. 
} In order to keep covariance manifest we choose to express $\dot\gamma_{ij}$ as $2 K_{ij}$.

In the following we adopt dimensional regularization in $d \equiv 4-\varepsilon$ dimensions and we work at lowest order in the coupling $\tilde m_4^2$. Then, at 1-loop, the only diagram contributing to the graviton propagator we need to evaluate is
\begin{align}\label{eq:diagram_loop}
\vcenter{\hbox{\includegraphics[width=0.3\textwidth]{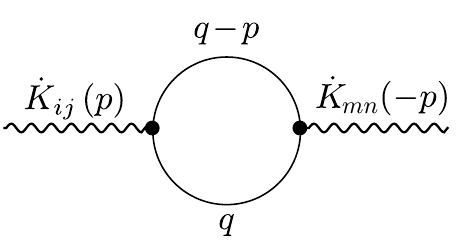}}} = i \Pi_{ijmn}(p) \frac{M_\text{Pl}^2}{2}\dot K_{ij}(p)\dot K_{mn}(-p)\;.
\end{align}
Indeed, tadpole diagrams with virtual massless fields vanish in dimensional regularization since they do not contribute to logarithmic divergences.
To maintain the correct dimensions, the scale $\Lambda_*$ is replaced by $\Lambda_{*d} = \Lambda_*\, \mu^{-\varepsilon /6}$, where $\mu$ is an arbitrary energy scale.\footnote{Note that in $d$ spacetime dimensions $\gamma^{(c)}_{ij}$ and $\pi^{(c)}$ have dimension $d/2-1$.} 
Additionally, the propagator for $\pi$  is 
\begin{align}\label{eq:diagram_propagator}
\vcenter{\hbox{\includegraphics[width=0.10\textwidth]{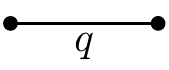}}} = \frac{-i}{-q_0^2 + c_s^2 \vect q^2 - i \epsilon} = \frac{-i}{\bar q^2 - i \epsilon} \;,
\end{align}
where in the last equality we have defined
\be
\bar q^\mu \equiv (q_0 , c_s \vect q) \;.
\ee

At this point we are ready to evaluate the amplitude of the diagram \eqref{eq:diagram_loop} as:
\be
\begin{split}
i \Pi_{ijmn}(p) & = \frac{1}{2} \times  \left(\frac{-4 i}{\Lambda_{*d}^3}\right)^2  \int \frac{\text d ^d q }{(2 \pi)^d} \frac{-i}{\bar q ^2 - i \epsilon} \frac{-i }{(\bar q - \bar p)^2 - i \epsilon}  \frac{1}{4}\left[q_i  (q-p)_j  + q_j (q-p)_i\right]\times \label{eq_selfE} \\ 
& \times \left[ q_m  (q-p)_n  + q_n (q-p)_m \right]\\
\end{split}
\ee
One can now insert a Feynman parameter $x$ and change the variable of integration to $k \equiv \bar q - \bar p x$. Notice that terms with powers of $\vect  p$ in the numerator are not relevant for gravitational waves and can be disregarded as they would generate operators containing $ \partial_i K_{ij}$ that vanish for transverse-traceless perturbations. The same also holds for terms proportional to the trace of the extrinsic curvature. Using the suffix ${}^{\rm (TT)}$ to denote that we restrict to these terms,  one gets
\be\label{eq_loop}
i \Pi_{ijmn}^{\rm (TT)}(p) = \frac{8  \mu^\varepsilon}{\Lambda_*^6c_s^{d+3}} \int_0^1 \text d x \int\frac{\text d ^d k}{(2 \pi)^d} \frac{  k_i k_j k_m k_n}{[k ^2 + \bar p^2 x(1-x) - i \epsilon]^2} \;.
\ee
Due to the rotational symmetry of the integral over $k$ in \eqref{eq_loop}, we can use
\begin{align}
k_i k_j k_m k_n & = \frac{\vect k^4}{d^2 -1}\left( \delta_{im}\delta_{jn} + \delta_{in}\delta_{jm} +\delta_{ij}\delta_{mn}\right)\;,
\end{align}
where the last term in the parenthesis can be dropped, since it yields a term proportional to the trace of the extrinsic curvature. 

After these steps, we define $\Delta \equiv \bar p^2 x (1-x)- i \epsilon$ and we compute  the integral in eq.~\eqref{eq_loop}  using that $\vect k^4= (k^2 -k_0^2)^2$ and 
\begin{align}
 \int \frac{\text d ^d k}{(2 \pi)^d} \frac{(k^2 -k_0^2)^2}{(k^2 + \Delta)^2}   =I\left[ 1 + \frac{\Omega_{d-1}}{\Omega_d}\int_0^\pi (\cos^4 \phi - 2\cos^2  \phi) \sin^{d-2}\phi \text d \phi  \right] = I \frac{d^2-1}{d(d+2)} \;,
\end{align}
where $I$ and $\Omega_d$,  the area of the $(d-1)$-sphere, are given by
\be
I \equiv \frac{i}{(4 \pi)^{d/2}} \frac{d(d+2)}{4} \frac{\Gamma( - d/2)}{\Delta^{-d/2}}\;,\qquad 
\Omega_d \equiv \frac{2 \pi^{d/2}}{\Gamma(d/2)} \;.
\ee
Thus, we get 
\begin{align}
i \Pi_{ijmn}^{\rm (TT)}(p) &=  \frac{2 i \mu^\varepsilon (\bar p^2 - i\epsilon)^{d/2}}{\Lambda_*^6c_s^{d+3}}  \frac{\Gamma(-d/2)}{(4 \pi)^{d/2}}  \frac{\Gamma(1+d/2)^2}{\Gamma(2+d)}  \, (\delta_{im}\delta_{jn} + \delta_{in}\delta_{jm})\; .
\end{align}
From this, by taking the limit $d \to 4 -\varepsilon$ and expanding at leading order in $\varepsilon$,  we obtain the divergent contribution to the effective action in momentum space,
\begin{align}\label{eq_1loopeffaction}
S_{\text{eff}} &=  \frac{M_\text{Pl}^2}{ 480 \pi^2\Lambda_*^6 c_s^7} \int \frac{\text d ^4 p\,}{(2\pi)^4}  \bar p^4 \dot K_{ij}(p) \dot K_{ij}(-p)\left[ \frac{1}{\varepsilon} + \frac{23}{15} - \frac{\gamma_{\text E}}{2} - \frac{1}{2} \log \left( \frac{\bar p^2}{ 4 \pi c_s^2 \mu^{2}}  - i \epsilon \right)  \right] \;,
\end{align}
where $\gamma_\text E$ is the Euler-Mascheroni constant.

After introducing the suitable counterterm to remove the divergent part in the limit $\varepsilon \to 0$, one is left with a dispersion relation for the gravitational waves of the form
\begin{equation}\label{eq_disprel}
\omega^2 = \vect k^2 - \frac{\vect k ^8(1-c_s^2)^2 }{480 \pi^2 \Lambda_*^6 c_s^7}\log \left( - (1-c_s^2) \frac{\vect k^2 }{\mu_0^{2}} - i \epsilon\right) \;.
\end{equation}
Here  $\mu_0$ is an unknown constant that must be fixed by experiments. This dispersion relation is not Lorentz-invariant and since the momenta relevant for observations are of order $\Lambda_3$, it is not compatible with the recent GW results (see \cite{Yunes:2016jcc,LIGO2017} for experimental constraints on GW modified dispersion relations) unless $\tilde m_4$ is very small or $c_s$ is very close to unity. Notice that the higher derivative correction cannot be set to zero since it runs logarithmically with the scale $\vect k^2$. Notice also that the correction to the propagation of GWs is there even when $c_s >1$ and the decay of the GW cannot take place. 
This is indeed consistent with the fact that the loop in \eqref{eq:diagram_loop} involves the propagation of off-shell $\pi$'s, hence there is no kinematic restriction to the calculation. 
If one starts with an action with $c_{\rm T}=1$, this condition is stable under radiative corrections. However, eq.~\eqref{eq_disprel} shows that higher-derivative non-Lorentz invariant operators are generated. 

In the calculation, we did not take into account loops of the Fadeev-Popov ghost fields. The ghosts appear in any theory with gauge redundancy as a way to express the determinant of the variation of the gauge condition with respect to the gauge parameters. In general, this determinant is field dependent and needs to be included in the action via the Fadeev-Popov procedure. This happens for the two gauges we are using in this paper, the Newtonian gauge and, in the appendix, the spatially flat one (in particular both gauges feature residual gauge freedom at zero momentum and this non-physical modes must be cancelled by the ghosts). The ghost action only depends on the chosen gauge condition and, in particular, it does not depend on the operators that describe the dynamics of the fluctuations around the FRW background. The loop of ghosts with two external graviton lines will therefore be independent of $\tilde m_4$ and as such not suppressed by the low scale $\Lambda_3$.

We also observe that unitarity of the S-matrix, in the form of the optical theorem, provides a non-trivial check of our results thus far. Indeed this theorem sets an equality between the imaginary part of the graviton self-energy (evaluated on-shell) and the decay rate times the energy we computed in \eqref{eq:decaygpp}. Furthermore, this relation can be expressed diagrammatically as
\begin{align}\label{eq:opticalth}
\vcenter{\hbox{\includegraphics[width=0.45\textwidth]{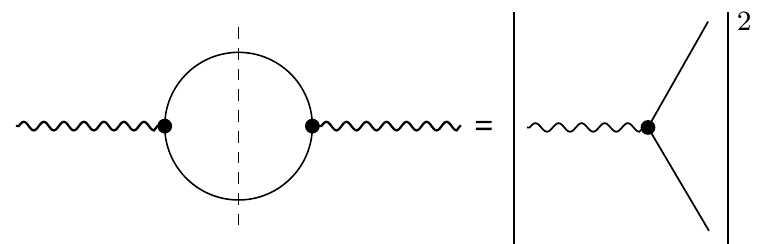}}}
\end{align}
and remarkably does not depend on the renormalization procedure one employs.

We can readily show this equality by evaluating the imaginary part of \eqref{eq_disprel} and comparing the result with $\Gamma \omega$. Of course the only term contributing to the imaginary part is the logarithm.  
For $c_s^2 <1$ the argument of the logarithm is negative so we have
\begin{equation}
\text{Im}\, \omega^2 = -\frac{\vect k^8(1-c_s^2)^2}{480 \pi^2 \Lambda_*^6 c_s^7}\text{Im} \log (-\vect k^2 (1-c_s^2) - i \epsilon) = \frac{\vect k^8(1-c_s^2)^2}{480 \pi \Lambda_*^6 c_s^7}  = \Gamma \omega.
\end{equation}
Conversely for $c_s^2 >1$ the argument of the logarithm is positive and as expected we find no imaginary part, in agreement with \eqref{eq:decaygpp} where in this case the result is zero.

The experimental constraints on the imaginary and real part of the dispersion relations are similar
\be
\frac{{\rm Im} \;\omega^2}{\omega^2} \lesssim \frac{1}{\omega d_S} \;,\qquad \frac{{\rm Re}\;(\omega^2 - k^2)}{\omega^2} \lesssim \frac{1}{\omega d_S} \;,
\ee
where $d_S$ is the distance of the source.
Indeed, neither the amplitude nor the phase of a given Fourier mode can have an order one modification travelling from the source to the detector.\footnote{For the real part the constraints come from comparing different frequencies, i.e.~looking at the distortion of the expected signal. In the case of a quadratic dispersion relation with $c_{\rm T} \neq 1$ the signal is not distorted and one has to rely on an optical counterpart, with somewhat looser bounds.} The bound reads \begin{equation}
 \frac{1}{\omega d_S}  \sim 10^{-18} \times\frac{2 \pi  \times100\, \text{Hz}}{\omega} \;\frac{40\, \text{Mpc}}{d_S}\;.
 \end{equation}

\subsection{\label{subsec:HD}Higher-derivative corrections}
The calculation above shows that radiative corrections generate operators suppressed by powers of $\partial/\Lambda_3$. Even if we concentrated on the logarithmic divergences, which do not depend on the UV physics, one expects that power divergences will be generated as well. These operators are in general not Lorentz invariant and affect the propagation of tensor modes: since $\partial/\Lambda_3$ is not very small in the recent observations of GWs, this setup is ruled out by observations. (We expect this conclusion to hold also in the case $c_s=1$, even though the calculable corrections of the previous Section vanish.) 
One has both a large decay rate of gravitational waves and a sizeable distortion of the signal.
It is however important to point out two possible ways out.

First, these conclusions do not apply to the operator $\delta g^{00} \delta K$ or, in the covariant language, to the cubic Galileon/Horndeski. Even though the strong coupling scale is $\Lambda_3$, it is easy to realise that the coupling with gravitational waves is very suppressed. Indeed if one considers this operator with a size that is relevant for modifications of gravity on cosmological scales (corresponding to a cutoff of order $\Lambda_3 $), one can easily read the coupling with gravity
\begin{equation}
H M_{\text{Pl}}^2 \delta g^{00} \delta K \sim H M_{\text{Pl}}^2 \dot \pi \partial_i\partial_j\pi \gamma_{ij} \;.
\end{equation}
We obtain a coupling that, if compared with the one of the previous sections, is suppressed by a much larger scale: 
\be
\Lambda_2 \equiv (H_0 M_{\text{Pl}})^{1/2}\;. 
\ee
This strongly suppresses both the decay rate of gravitational waves and the loop corrections to the propagation. 

Since this point is quite important, it is worthwhile repeating it in the covariant language. Making explicit the dependence on the scales $\Lambda_2$ and $\Lambda_3$ following \cite{Pirtskhalava:2015nla}, one starts with a Horndeski action of the form
\be
\label{WBG}
L_2 = \Lambda_2^4 G_2\left(\frac{(\partial\phi)^2}{\Lambda_2^4}\right)\;, \qquad L_3 = \Lambda_2^4 G_3\left(\frac{(\partial\phi)^2}{\Lambda_2^4}\right)  \frac{\Box\phi}{\Lambda_3^3}\;,\qquad  L_{4,5} \ldots \;,
\ee
where the explicit form of the $L_4$ and $L_5$ operators is given in App.~\ref{App:Horndeski}. 
The action is characterised by the two scales $\Lambda_2  \gg \Lambda_3$. This form of the action is stable under radiative corrections: the functions $G_i$ receive corrections that are parametrically suppressed by $(\Lambda_3/\Lambda_2)^4 \ll 1$ compared to original action. (In particular this implies the stability of the condition that GWs travel luminally, at the 2-derivative level, $c_\text{T}=1$ \cite{Creminelli:2017sry}.) This result is based on the non-renormalisation theorem of Galileons \cite{Luty:2003vm} and on the small breaking induced by gravity. The gravitational interactions present in the covariant Galileons \cite{Deffayet:2009wt} are of the form
\begin{align}\label{fig:covGal}
\vcenter{\hbox{\includegraphics[width=0.22\textwidth]{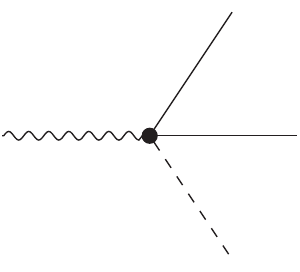}}} \quad\vcenter{\hbox{\includegraphics[width=0.22\textwidth]{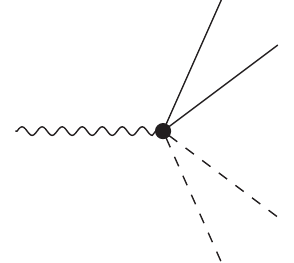}}} \quad\vcenter{\hbox{\includegraphics[width=0.22\textwidth]{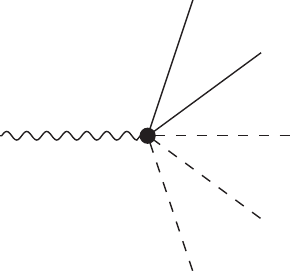}}}
\end{align}
Solid lines represent a single derivative acting on the scalar, while dashed lines more than one derivative; the wavy line is a graviton. The crucial point is that one has one graviton and thus one power of $1/M_{\text{Pl}}$ for each $(\partial\phi)^2$. This motivates the scaling of eq.~\eqref{WBG} for the  Horndeski Lagrangian and shows that the operators are renormalized in a very suppressed way \cite{Pirtskhalava:2015nla}. 

Let us now consider the renormalisation of operators with external graviton lines (we take the graviton canonically normalized, i.e.~a dimension one field $\gamma^{(c)}$):
\begin{align}\label{}
\vcenter{\hbox{\includegraphics[width=0.2\textwidth]{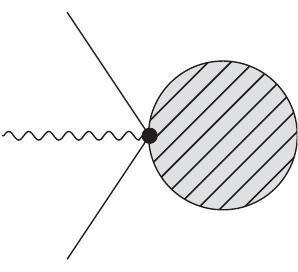}}}
\end{align}
The scaling still works in the same way: one power of $1/M_{\text{Pl}}$ for each $(\partial\phi)^2$ (one can have more powers of $(\partial\phi)^2$ as external legs, but each carries its $1/M_{\text{Pl}}$ since the functions $G_i$ above are characterized by $\Lambda_2$ and not $\Lambda_3$ in the action). This implies the generation of operators of the schematic form
\be
\Lambda_3^4 F\left(\frac{(\partial\phi)^2}{\Lambda_2^4}, \frac{\partial^2\phi}{\Lambda_3^3},\frac{\gamma^{(c)}}{\Lambda_3}, \frac{\partial}{\Lambda_3}\right) \;. 
\ee
Since on the background solution one has $\dot\phi^2 \sim \Lambda_2^4$ one sees that the action for tensors is characterised by the only scale $\Lambda_3$: in particular one has sizeable corrections to the propagation if the frequency is not well below $\Lambda_3$. Notice however that the conclusion does not apply to the cubic Horndeski, i.e.~to the first interaction of \eqref{fig:covGal}. Since graphs must be 1PI one is forced to have one leg with a single derivative inside the loop: this changes the scaling and suppresses the final result by $(\Lambda_3/\Lambda_2)^4$. 
Therefore a theory with only cubic Galileon/Horndeski is viable since it does not affect the graviton propagation at the scale $\Lambda_3$ (or at least it is technically natural to make this assumption). Notice that this setup is consistent since the non-renormalization theorem guarantees that if quartic and quintic terms are zero at the beginning, they will be generated only with a very suppressed coefficient. One can generalise the argument to beyond Horndeski theories following \cite{Santoni:2018rrx}. 

The second caveat is that the theory that describes cosmological perturbations may break down at energies parametrically lower than $\Lambda_3$ \cite{deRham:2018red}. Of course nothing forbids that a theory changes before reaching its unitarity cutoff. In this case the EFT of DE cannot be used to describe the recent observations of propagation of gravitational waves. Since we do not know of any explicit UV completion of the DE theories we are studying, it is difficult to reach general conclusions. Naively, one expects the speed of gravitons to approach the speed of light as $c_{\textrm{T}} - 1 \sim M^2 /\omega^2 $, where $M$ is the typical mass of the new degrees of freedom. Therefore, to satisfy the experimental bounds one needs $M \lesssim (10^{11} \, \rm{km})^{-1}$.
It would be interesting to study the constraints imposed by causality and analyticity on a scenario in which gravitational waves have a different speed at different frequencies. In the analogous problem of light propagating in a material, one can derive general conclusions on the absorption of light given its frequency-dependent speed. Indeed the real and imaginary part of the index of refraction are related by the Kramers-Kronig relations. It is worthwhile studying whether similar techniques can be applied to the propagation of gravitational waves.

\section{\label{sec:conclusions}Conclusions}

The observation of gravitational waves has opened a new way of constraining dark energy and modified gravity.  This is made possible by the fact that the cutoff of the scalar-tensor EFT describing  dark energy, $\Lambda_3 = (\MP H_0^2)^{1/3}$, lays within the LIGO/Virgo band.  
At these energy scales, interactions involving gravitons and dark energy fluctuations become large. In the presence of  spontaneous breaking of Lorentz invariance,  this makes gravitons decay at a  catastrophically large rate.

As explained in Sec.~\ref{sec:EFTDE},  to be compatible with the GW170817 measurements
we have restricted our study to theories where gravitons propagate at the speed of light. In the covariant language (see App.~\ref{App:Horndeski}), these are described by the Lagrangian (see e.g.~\cite{Creminelli:2017sry})
\be
\label{eq:BHredux}
 L_{c_T=1}=     P(\phi,X) + Q(\phi,X) \Box\phi+ f(\phi,X) \, {}^{(4)}\! R - \frac4X f_{,X} (\phi,X) (\phi^{;\mu} \phi^{;\nu} \phi_{;\mu \nu} \Box \phi - \phi^{;\mu} \phi_{;\mu \nu} \phi_{;\lambda} \phi^{;\lambda \nu}) \;,
\ee
where $X \equiv g^{\mu \nu} \partial_\mu \phi \partial_\nu \phi$.
Moreover, we have focused  on cubic interactions.  In this case, two channels of decay are possible: $\gamma \to \pi \pi$ from the  coupling $ \alpha_{\rm H} \MP^2 \ddot \gamma_{ij} \partial_i \pi \partial_j \pi$ and $\gamma \to \gamma \pi$ from the coupling $ \alpha_{\rm H}\MP^2  \dot \pi \dot \gamma_{ij}^2$. Here $\alpha_{\rm H}$ is a dimensionless time-dependent function measuring the beyond Horndeski character of the theory \cite{Gleyzes:2014dya,Gleyzes:2014qga}. It is defined in eq.~\eqref{alphaH} and for the theory \eqref{eq:BHredux} above is given by
$\alpha_{\rm H} = - {2X f_{,X}}/{f} $.

We have studied the decay rate for $\gamma \to \pi \pi$ in Sec.~\ref{sec:decayrate}, finding that it is roughly given by $\Gamma_{\gamma \to \pi \pi }\sim \alpha_{\rm H}^2 \omega^7/\Lambda_{3}^6$ (the full expression can be found in eq.~\eqref{eq:decaygpp} with \eqref{Lambdadef}). The coupling $\gamma \gamma \pi$ contains one less derivative than $\gamma \pi \pi$. Therefore,
the decay rate for $\gamma \to \gamma \pi$ 
is much smaller: $\Gamma_{\gamma \to \gamma \pi }\sim \alpha_{\rm H}^2 \omega^5/\Lambda_{2}^4$, where $\Lambda_2 = (\MP H_0)^{1/2}$. This decay gives constraints on $\alpha_{\rm H}$ that are much looser than the other channel and for this reason it is studied in App.~\ref{app:gamma2pi}.  

The absence of this effect at LIGO/Virgo frequencies $\omega \sim \Lambda_3$ implies that $\alpha_{\rm H}$ is practically zero. 
Thus, the surviving theory is
\be
\label{surviving1}
L_{c_T=1, \text{ no decay}}=   f(\phi) \, {}^{(4)}\! R +  P(\phi,X) + Q(\phi,X) \Box\phi \;.
\ee
It is interesting to formulate this theory in the context of the more general Degenerate Higher-Order Scalar-Tensor (DHOST) \cite{Langlois:2015cwa}  or Extended Scalar-Tensor \cite{Crisostomi:2016czh} theories. These theories can be obtained starting from beyond Horndeski and performing an invertible conformal transformation that depends on $X$, i.e.~$g_{\mu \nu} \to C(\phi,X) g_{\mu \nu}$ \cite{Zumalacarregui:2013pma,Crisostomi:2016czh,Achour:2016rkg}  (we assume $C$ is not linear in $X$  so that the transformation is invertible). Since this does not change the  light-cone, we can do the same with the theory above, obtaining 
\be
 L_{c_T=1, \text{ no decay}}=   P(\phi,X) +  Q(\phi,X) \Box\phi +  C (\phi,X) {}^{(4)}\! R +\frac{6 C_{,X}(\phi,X){}^2}{C(\phi,X)}  \phi^{;\mu} \phi_{;\mu \nu} \phi_{;\lambda} \phi^{;\lambda \nu} \;. \label{DHOST}
\ee
Here we have redefined the free functions $P$ and $Q$ after the transformation and reabsorbed the dependence on $f(\phi)$ in $C$.
This is the most general degenerate theory  compatible with $c_\text{T}^2=1$ and with the absence of graviton decay.\footnote{In terms of the dimensionless  coefficients defined in Ref.~\cite{Langlois:2017mxy}, for DHOST theories we find that neither $\alpha_{\rm H}$ nor $\beta_1$, the coefficient  parameterizing the presence of higher-order
operators, vanish. However, in the absence of decay these coefficients are not independent but are related by $\alpha_{\rm H}= - 2\beta_1$. This implies that the screening mechanism based on quartic terms studied in \cite{Crisostomi:2017lbg,Langlois:2017dyl,Dima:2017pwp} is absent. We thank M.~Crisostomi and K.~Koyama for pointing this out to us.}

The closeness of the cutoff to the LIGO/Virgo band is also responsible for modifying the dispersion relation of the gravitational waves, see Sec.~\ref{sec:loops}.
For $c_s <1$, this is expected because the dispersion is related to the decay by the optical theorem, as we explain in Sec.~\ref{sec:4.1}. For $c_s>1$, even if the decay of gravitons is kinematically forbidden, the loop corrections to the graviton propagation are still present and give practically the same bound. In the case $c_s=1$, the decay rate and the calculable part of the loop corrections that we studied vanish. On the other hand, power-law divergent terms are expected and would provide similar constraints to those obtained for $c_s \neq 1$. We conclude that the absence of $\tilde m_4^2$ holds for any value of the scalar speed $c_s$.
Interestingly, as explained in Sec.~\ref{subsec:HD} radiative corrections in the surviving theory \eqref{surviving1} (and its degenerate version \eqref{DHOST}) do not generate measurable effects in the graviton dispersion relation even at the LIGO/Virgo scales. 

As already mentioned in the article, our conclusions do not hold if the  theories at hand break down at a scale parametrically smaller than $\Lambda_3$ \cite{deRham:2018red}. It would be interesting to investigate further whether an example of such a proposal  can be constructed that successfully reproduces  GR on short scales. Moreover, we stress again that in this article we have studied the perturbative decay of gravitational waves, neglecting possible coherent effects. Given the very high occupation number of gravitons in the observed waves, we expect that these effects are indeed important and that their absence can be used to rule out another corner of the parameter space of these theories. We leave this investigation for the future. Another interesting direction is to study whether these effects are also relevant for theories that are not described by the EFT of DE.

\section*{Acknowledgements}
It is a pleasure to thank D. Pirtskhalava, L.~Santoni, L.~Senatore, E.~Trincherini and G.~Villadoro for interesting discussions and E.~Bellini, M.~Crisostomi, G.~Cusin, J.~Ezquiaga, P.~Ferreira, K.~Koyama, M.~Lagos, S.~Melville, J.~Noller, J.~Sakstein and M.~Zumalacarregui for useful comments on the draft. M.~L.~acknowledges financial support from the Enhanced Eurotalents fellowship, a Marie Sklodowska-Curie Actions Programme.
\vspace{0.3cm}

\appendix

\section{Connection with Horndeski and beyond Horndeski theories}
\label{App:Horndeski}
\newcommand{\Gtwo}{G_2{}}
\newcommand{\Gthree}{G_3{}}
\newcommand{\Gfour}{G_4{}}
\newcommand{\Gfive}{G_5{}}
\newcommand{\Ftwo}{F_2{}}
\newcommand{\Fthree}{F_3{}}
\newcommand{\Ffour}{F_4{}}
\newcommand{\Ffive}{F_5{}}
\newcommand{\Hfour}{F_4{}}
\newcommand{\Hfive}{F_5{}}

Following \cite{Creminelli:2017sry}, in this appendix we connect the EFT action \eqref{total_action} with the covariant formulation of Horndeski \cite{Horndeski:1974wa,Deffayet:2011gz} and beyond Horndeski \cite{Zumalacarregui:2013pma} theories {\em \`a la} GLPV \cite{Gleyzes:2014dya,Gleyzes:2014qga}. In particular, let us consider a scalar field $\phi$ 
and define $X\equiv g^{\mu \nu} \partial_\mu \phi \partial_\nu \phi$ and  $\Box \phi \equiv \phi^{;\mu}_{;\mu}$. The symbol  $\epsilon_{\mu \nu \rho \sigma }$ denotes the totally antisymmetric Levi-Civita tensor, a comma  a partial derivative with respect to the argument and a semicolon  the covariant derivative. 

The scalar field dynamics of GLPV theories is governed by the action  \cite{Gleyzes:2014dya,Gleyzes:2014qga}
\be
S= \int d^4 x \sqrt{-g} \sum_I L_I\;,
\ee
where 
\be
\begin{split}
L_2 & \equiv  \Gtwo(\phi,X)\;,   \qquad L_3  \equiv  \Gthree(\phi, X) \, \Box \phi \;,  \\
L_4 & \equiv \Gfour(\phi,X) \, {}^{(4)}\!R - 2 \Gfour_{,X}(\phi,X) (\Box \phi^2 - \phi^{; \mu \nu} \phi_{ ; \mu \nu}) \\
& -\Hfour(\phi,X) \epsilon^{\mu\nu\rho}_{\ \ \ \ \sigma}\, \epsilon^{\mu'\nu'\rho'\sigma}\phi_{; \mu}\phi_{; \mu'}\phi_{; \nu\nu'}\phi_{; \rho\rho'}\;, \\
L_5 & \equiv \Gfive(\phi,X) \, {}^{(4)}\!G_{\mu \nu} \phi^{; \mu \nu} \\ 
& +  \frac13  \Gfive_{,X} (\phi,X) (\Box \phi^3 - 3 \, \Box \phi \, \phi_{; \mu \nu}\phi^{; \mu \nu} + 2 \, \phi_{; \mu \nu}  \phi^{; \mu \sigma} \phi^{; \nu}_{\ ; \sigma})  \\ \;&
 \quad- \Hfive (\phi,X) \epsilon^{\mu\nu\rho\sigma}\epsilon^{\mu'\nu'\rho'\sigma'}\phi_{; \mu}\phi_{; \mu'}\phi_{; \nu \nu'}\phi_{; \rho\rho'}\phi_{; \sigma\sigma'} \label{LHBH}\,,
\end{split}
\ee
and  $F_4$ and $F_5$ are related by
\be
\label{deg}
X G_{5,X} F_4 = 3 F_5 \left[ G_4 - 2 X G_{4,X} - (X/2) G_{5,\phi} \right] \;,
\ee
in order for the theory to be degenerate \cite{Langlois:2015cwa}.

The relevant  parameters in eq.~\eqref{total_action} can be written in terms of the covariant functions $G_4$, $G_5$, $F_4$ and $F_5$ above. We find 
\be
\begin{split}
M^2 & \equiv M_*^2 f +2 m_4^2 =  2 G_4 - 4 X G_{4,X} - X \big( G_{5,\phi} + 2 H \dot \phi  G_{5,X}  \big) 
+ 2 X^2 F_4 -6H \dot \phi X^2 F_5\;, \\
m_4^2 & = \tilde m_4^2 + X^2 F_4 - 3 H \dot \phi X^2 F_5   \;, \\
\tilde m_4^2 & = - \big[ 2 X G_{4,X} + X G_{5,\phi} + \big(H \dot \phi - \ddot \phi \big) X G_{5,X}  \big] \;, \\
m_5^2 & = X \big[ 2 G_{4,X} +   4 X G_{4,XX} + H \dot \phi  ( 3 G_{5,X}  + 2  X  G_{5,XX})  + G_{5,\phi}  \\ & + X G_{5,X \phi}   - 4 X F_4 -2X^2 F_{4,X}  + H \dot \phi  X \big( 15 F_5 + 6 X F_{5,X} \big) \big] \;,\\
m_6  & = \tilde m_6 -   3 \dot \phi X^2 F_{5}      \;, \\
 \tilde m_6  &= -  \dot \phi X  G_{5,X}  \;, \\
 m_7 & = \frac{1}{2} \dot \phi X \big(3 G_{5,X}+2 X G_{5,XX} +15 X F_{5} + 6 X^2 F_{5,X} \big) \;. \label{alphasGF}\end{split}
\ee
For theories with luminal gravitational waves we have $G_{5,X}=0=F_5$ and $2 G_{4,X}-X F_4+G_{5, \phi}=0$ (see eqs.~\eqref{m4zero} and \eqref{mpercT1} and e.g.~\cite{Creminelli:2017sry}).

\section{Generic disformal frame}
\label{App:Frame}

In Sec.~\ref{subsec:decayrate} we have seen that the parameter $\tilde m_4^2$ has to vanish to suppress the gravitational wave decay. We have made the calculation in a frame 
where  gravitons travel at a speed $c_\text{T}=1$, so that several of the EFT parameters are absent from the beginning.  
Combining the new constraint $\tilde m_4^2=0$ with  those coming from the  speed of gravitons, eqs.~\eqref{m4zero} and \eqref{mpercT1}, one finds that the EFT simplifies considerably:
\be
\dot f = m_4^2 = \tilde m_4^2= m_5^2 = m_6 = \tilde m_6 = m_7 =0  \;,
\ee 
where the time independence of $f$ can be set by a conformal transformation.
Here we want to see the consequences of the absence of gravitational wave decay in a generic disformal frame and show that our results can be written in a frame independent way. 

Exceptionally in this appendix we will use the following notation: we will denote by a hat quantities in the special frame where $\hat c_\text{T}=1$, while quantities without a hat are in a generic frame. Starting from the generic action \eqref{total_action}, it is possible to show that the cubic interaction $\gamma \pi \pi$ is controlled by the scale 
\be
\label{Lambdatre}
\Lambda_*^3 = \frac{2 \sqrt{2}M H^2 c_\text{T}^2 \alpha}{ \left[1 + \alpha_{\rm H} - c_\text{T}^2 (1+\alpha_{\rm V}) \right] (1+\alpha_{\rm H})} \;,
\ee
where the dimensionless quantity
\be
\alpha \equiv \frac{2 c+4 m_2^4}{M^2 H^2} + \frac32 \left( \frac{M_*^2 \dot f - m_3^3}{M^2 H} \right)^2 \;
\ee
sets the normalization of the scalar fluctuations \cite{Gleyzes:2014rba} and we have also defined the dimensionless quantities (see e.g.~\cite{Dima:2017pwp})
\be
c_\text{T}^2 \equiv 1 -\frac{2 m_4^2}{M^2} \;, \qquad \alpha_{\rm H} \equiv \frac{2 (\tilde m_4^2 - m_4^2)}{M^2} \;, \qquad \alpha_{\rm V} \equiv - \frac{2 m_5^2}{M^2} \;.
\ee

Generalizing the calculations of Sec.~\ref{subsec:decayrate} in the frame where $c_\text{T} \neq 1$ (in this frame photons and gravitons move at the same speed, as required by experiments, but not equal to unity) we can derive the decay rate in a generic frame. This reads
\be
\Gamma_{\gamma \to \pi \pi} = \frac{ E_{p}^7 \left( 1 -  c_s^2/  c_\text{T}^2 \right)^2}{480 \pi   c_s^7   \Lambda_*^6} \;,
\ee
where $E_{ p} =   c_\text{T}   p$ and $\Lambda_*^6$ is obtained from squaring eq.~\eqref{Lambdatre} above. This expression generalizes the one in eq.~\eqref{eq:decaygpp} to a generic frame.

We can now check that this result can be obtained from the decay rate in the frame with $\hat c_\text{T}= 1$, i.e.~(see eq.~\eqref{eq:decaygpp}) 
\be
\label{Gammahat}
\hat \Gamma_{\gamma \to \pi \pi} = \frac{ {\hat p}^7 \left( 1 -  \hat c_s^2 \right)^2}{480 \pi \hat  c_s^7 \hat  \Lambda_*^6} \;, 
\ee
with (see eq.~\eqref{Lambdadef})
\be
\label{Lambdatrehat}
\hat \Lambda_*^3 = \frac{\sqrt{2} \hat M_{\rm Pl} \hat H^2 \hat \alpha}{\hat \alpha_{\rm H} (1+ \hat\alpha_{\rm H})} \;.
\ee
(Notice that in the $\hat c_\text{T}=1$ frame $\hat \alpha_{\rm V} = - \hat \alpha_{\rm H}$ and one recovers this equation from eq.~\eqref{Lambdatre}.)
When moving from the $\hat c_\text{T}= 1$ to the $c_\text{T} \neq 1$ frame, momenta do not change (i.e. $\hat p = p$) but the scale $\hat \Lambda_*^6$ gets rescaled.
Indeed, using the effects of a disformal transformation studied in \cite{Gleyzes:2015pma,DAmico:2016ntq,Langlois:2017mxy} one can show that 
 \be
\hat M_{\rm Pl} = c_\text{T}^{1/2} M \;, \qquad \hat H = H/c_\text{T}\;,
\ee
and
\be
\hat \alpha = \frac{4 \alpha c_\text{T}^4}{\left[ 1 + \alpha_{\rm H} + c_\text{T}^2 (1+\alpha_{\rm V}) \right]^2}\;, \qquad \hat \alpha_{\rm H} = \frac{1 + \alpha_{\rm H} - c_\text{T}^2 (1+\alpha_{\rm V})}{1 + \alpha_{\rm H} + c_\text{T}^2 (1+\alpha_{\rm V})} \;.
\ee
Confronting eqs.~\eqref{Lambdatre} and \eqref{Lambdatrehat} using these expressions shows that 
$\hat \Lambda_*^3 = c_\text{T}^{1/2} \Lambda_*^3$.
Using this result and $\hat c_s = c_s/c_\text{T}$ in eq.~\eqref{Gammahat} one sees that the dimensionless  decay rate $\Gamma/H$ is invariant,
\be
\frac{\hat \Gamma_{\gamma \to \pi \pi}}{\hat H} = \frac{\Gamma_{\gamma \to \pi \pi}}{H} \;,
\ee
as expected.

To conclude, eq.~\eqref{Lambdatre} shows that the frame-invariant combination of parameters that is constrained by the absence of decay is
\be
\label{ottootto}
\frac{1+\alpha_{\rm H} - c_\text{T}^2 (1+\alpha_{\rm V})}{2} = \frac1{M^2} \left[ \tilde m_4^2 + m_5^2 \left(1 - \frac{2  m_4^2}{M^2} \right) \right] =0 \;.
\ee
From eq.~\eqref{alphasGF}, for a quartic GLPV theory this constraint reads
\be
2 G_{4,X}^2 - X G_{4,X} F_4 + 2 G_4 G_{4,XX}-2 G_4 F_4 -X F_{4,X} G_4=0 \;.
\ee
As expected, eq.~\eqref{ottootto} cannot be put to zero by a disformal transformation $g_{\mu \nu} \to g_{\mu \nu}  + D(\phi,X) \phi_{; \mu} \phi_{; \nu}$, as one can check using that
\be
\alpha_{\rm H} \to \frac{1+\alpha_{\rm H}}{1+\alpha_{\rm X}} - 1 \;, \qquad \alpha_{\rm V} \to \frac{1+\alpha_{\rm V}}{(1+\alpha_{\rm D})(1+\alpha_{\rm X})} - 1 \;, \qquad c_{\rm T}^2 \to c_{\rm T}^2 (1+\alpha_{\rm D}) \;,
\ee
where $\alpha_{\rm D} \equiv - XD/(1+XD)$ and $\alpha_{\rm X}\equiv-X^2 D_{,X}$.

\section{\label{app:flatgauge}Interactions in spatially-flat gauge}

\subsection{Gauge transformation} \label{gaugetransfsec}

To write the metric in Newtonian gauge, we start with the general decomposition
\be
\text ds^2 = - ( 1 + 2 \Phi) \text dt^2 + a(t)^2 \left( \left( e^\gamma \right)_{ij} - 2 \Psi \delta_{ij} \right) \left( \text dx^i + N^i \text dt \right) \left( \text dx^j + N^j \text dt \right) \ ,
\ee
where $\delta^{ij} \gamma_{ij} = 0$.  We can further decompose the vector part $N^i$ into a scalar and a transverse vector as $N_i = \partial_i \psi + \hat N_i$ where $\partial_i \hat N_i =0$ (in this section, indices are raised and lowered using the unperturbed metric $\bar g_{00} = -1$ and $\bar g_{ij} = a(t)^2 \delta_{ij}$, we use $\partial^2 \equiv \partial_i \partial_i$, and hatted quantities are divergenceless, unrelated to the change of frame in App.~\ref{App:Frame}).  To go to Newtonian gauge, we use three diffeomorphisms to make the tensor transverse, $\partial_j \gamma_{ij} = 0$, and one diffeomorphism to make the vector transverse, $\psi = 0$.  In this gauge, we also have the Goldstone mode $\pi ( x )$ which appears explicitly in the action (i.e. after the Stueckelberg trick).   

Another common gauge choice is the spatially-flat gauge (see e.g.~\cite{Malik:2008im}), where the metric is written in the general decomposition 
\be
\text ds^2 = - ( 1 + \delta N )^2 \text d \tilde t^2 + a(\tilde t )^2 \left( e^{\tilde \gamma} \right)_{ij} \left( \text d \tilde x^i + \tilde N^i \text d \tilde t \right) \left( \text d \tilde x^j + \tilde N^j \text d \tilde t \right) \ .
\ee
The four gauge conditions in this case are that the tensor is transverse and traceless, $\tilde \partial_j \tilde \gamma_{ij} = 0$ and $\delta^{ij} \tilde \gamma_{ij} = 0$.  Thus, in the decomposition of the vector, $\tilde N_i = \tilde \partial_i \tilde \psi + \hat{ \tilde N}_i$, the scalar $\tilde \psi$ is still present (here $\tilde \partial_\mu \equiv \partial / \partial \tilde x^\mu$).  The Goldstone field in this gauge is denoted $\tilde \pi ( \tilde x )$.  

Now, we wish to find the gauge transformation that connects the two above gauges to linear order.  Under the gauge transformation $x^\mu \rightarrow \tilde x^\mu = x^\mu + \xi^\mu$, the metric changes as normal
\be
\tilde g_{\mu \nu} ( \tilde x ( x ) ) = g_{\rho \sigma} ( x ) \frac{\partial x^\rho}{\partial \tilde x^\mu} \frac{ \partial x^\sigma}{\partial \tilde x^\nu} \ . 
\ee
Infinitesimally, this gives 
\be
\Delta g_{\mu \nu} ( x ) \equiv \tilde g_{\mu \nu} ( x ) - g_{\mu \nu} ( x ) = - \xi^\sigma \partial_\sigma g_{\mu \nu} - \partial_\mu \xi_\nu - \partial_\nu \xi_\mu 
\ee
where on the right-hand side, and in the rest of this section, all derivatives without a tilde are taken with respect to the $x$ coordinates, and all fields are evaluated at the point $x$.  Expanding the metric around a time-dependent background $g_{\mu \nu} ( x ) = \bar g_{\mu \nu} ( t ) + \delta g_{\mu \nu} ( x )$, this gives the following relation between the fluctuations
\be
\delta \tilde g_{\mu \nu} ( x ) = \delta g_{\mu \nu} ( x ) + \Delta g_{\mu \nu} ( x ) \ . 
\ee
Additionally, the transformation of the Goldstone field is dictated by the fact that it non-linearly realizes time-diffemorphisms: $\tilde \pi ( \tilde x ( x ) ) = \pi ( x ) - \xi^0 ( x )$, or infinitessimally as 
\be
\Delta \pi ( x ) \equiv \tilde \pi ( x ) - \pi ( x ) = - \xi^\sigma \partial_\sigma \pi - \xi^0 \ . 
\ee
This gives the following relationships among the fields
\begin{align}
\delta N & = \Phi + \partial_0 \xi_0 \\
a^2  \tilde \gamma_{ij} & = a^2 \left( \gamma_{ij} - 2 \Psi \delta_{ij} -2 H \xi^0 \delta_{ij} \right)-  \partial_i \xi_j - \partial_j \xi_i  \label{hijeqtrans} \\
  \tilde N_i & =  N_i - \partial_0 \xi_i - \partial_i \xi_0 \label{nieq}  \\
 \tilde \pi & = \pi - \xi^0 \ . \label{pitransf} 
\end{align}

It is also convenient to parametrize the spatial part of the diffeomorphism into a scalar and a transverse vector: $\xi_i = \partial_i \xi + \hat \xi_i$ where $\partial_i \hat \xi^i = 0$.  Requiring that both $\gamma_{ij}$ and $\tilde \gamma_{ij}$ be transverse and traceless gives
\be \label{xizeroandxi}
\xi^0 = - \frac{\Psi}{H} \ , \quad \text{and} \quad  \partial^2 \xi = 0 \ . 
\ee
The remaining tensor part of \eqn{hijeqtrans} gives 
\be
\partial_i \hat \xi_j + \partial_j \hat \xi_i = a^2 ( \gamma_{ij} - \tilde \gamma_{ij} ) \ , 
\ee
while the scalar and vector parts of \eqn{nieq} give 
\begin{align}
\partial^2 \xi_0  = - \partial^2 \tilde \psi \ , \quad \text{and} \quad  \partial_0 \hat \xi_i  = \hat N_i - \hat{\tilde N}_i \ . \label{apples}
\end{align}

%

\subsection{Vertices in spatially-flat gauge}

In this subsection we redo the computations of \secref{sec:EFTDE} in the spatially-flat gauge.  As we will see, because we are in a different gauge, the relevant vertices emerge from different terms in the action.  In the spatially-flat gauge, $\delta N$ and $\tilde N^i$ are Lagrange multipliers, and for the cubic action, we only need their expressions to first order \cite{Maldacena:2002vr}.  Variation of the action with respect to $\tilde \psi$ gives the constraint equation for $\delta N$, and variation with respect to $\delta N$ gives the constraint equation for $\tilde \psi$:
\begin{align}
\begin{split}
\frac{\delta S }{\delta \tilde \psi} & =  a^3 \partial^2  \left( - 2 M_{\text{Pl}}^2 \left( \dot H \tilde \pi + H \delta N \right) +  m_3^3 \left( \delta N - \dot{ \tilde \pi} \right) \right)  \\
\frac{\delta S }{\delta \,  \delta N} & = 2 a^3 \left(  2 m_2^4 + 3 H m_3^3 - M_{\text{Pl}}^2 ( 3 H^2  + \dot H ) \right) \delta N + a^3 \left( 2 M_{\text{Pl}}^2 \dot H - 4 m_2^4 - 3 H m_3^3 \right) \dot{ \tilde \pi }\\
& \quad + 3 a^3 \dot H \left( m_3^3 - 2 H M_{\text{Pl}}^2 \right) \tilde \pi + a \partial^2 \left(  \left( m_3^3 + 4 H \tilde m_4^2 \right) \tilde \pi + \left( m_3^3 - 2 H M_{\text{Pl}}^2 \right) \tilde \psi   \right)  \ . 
\end{split}
\end{align}  
In the high-energy ($H \tilde \pi \ll \dot{ \tilde \pi }$) and sub-horizon ($H \ll \partial$) limits, setting the above to zero gives the solutions (see \cite{Cheung:2007sv} for the case $\tilde m_4^2 =0$),
\be
\delta N = \frac{ m_3^3}{ m_3^3 - 2 M_{\text{Pl}}^2 H} \dot{\tilde \pi} \ , \quad \text{and} \quad \tilde \psi = - \frac{ m_3^3 + 4 H \tilde m_4^2}{m_3^3 - 2 H M_{\text{Pl}}^2} \tilde \pi \ . 
\ee
Notice that one can also obtain the same results by directly using the equations for the gauge transformation in  \secref{gaugetransfsec}.   These solutions can then be plugged back into the action so that it is simply a functional of $\tilde \gamma_{ij}$ and $\tilde \pi$.  

As before, one can then look at the quadratic Lagrangian to find the canonical normalization of the fields and the speed of sound for $\tilde \pi$.  Because we have not changed the tensor part of the metric, the normalization for $\tilde \gamma_{ij}$ is the same as in \eqn{eq:gammacan}.  For $\tilde \pi$, we can use the results of the last section to immediately see the answer.  Using eqs.~\eqref{pitransf}, \eqref{xizeroandxi}, \eqref{phisol}, \eqref{Phitopidot} and \eqref{eq:pican}, we find
\be \label{changeinpi}
\tilde \pi = \frac{ 2 H M_{\text{Pl}}^2 - m_3^3 }{4 H \tilde m_4^2 + 2 H M_{\text{Pl}}^2} \pi  =  \frac{ 2 H M_{\text{Pl}}^2 - m_3^3}{\sqrt{2} H M_{\text{Pl}} (3 m_3^6 + 4 M_{\text{Pl}}^2 ( c + 2 m_2^4) )^\frac{1}{2}}  \pi^{(c)} \ .
\ee
Because only the normalization of $\pi$ changes between the gauges, the speed of sound $c_s^2$ is the same as in \eqn{csquared}.  

Now we move on to the non-linear $\gamma \pi \pi$ vertex.  This vertex receives contributions both from the Einstein-Hilbert term $ S_{\text{EH}}$ in \eqn{eq:EHaction} and the dark-energy term $ S_4$ in \eqn{eq:term4}.  There are two different contributions from $N^{-1} \left( E_{ij} E^{ij} -E^2 \right)$ in the Einstein-Hilbert term \eqn{eq:EHaction}:  the first has the form $\delta N \dot{\tilde{\gamma}}_{ij} \partial_i \partial_j \tilde \psi$ and comes from the $E_{ij}E^{ij}$ term, and the second has the form $\partial_i \tilde \psi \, \partial_j \tilde \psi \, \partial^2  \tilde \gamma_{ij}$ and comes from both $E_{ij} E^{ij}$ and $E^2$.  For the last term mentioned, we can use the linear equation of motion $ \partial^2 \gamma_{ij} = a^{2}  \ddot \gamma_{ij}$ so that the vertex has two time and two spatial derivatives, which is the form in \eqn{gammapipi}.  More specifically, we have
\begin{align}
 S_{\text{EH}}  & \supset   \frac{1}{2} M_{\text{Pl}}^2  \int \text d^4 x \,  a \left(  \alpha_{ N} \alpha_\psi  \,   \dot{\tilde \pi}  \dot{\tilde \gamma}_{ij} \partial_i \partial_j \tilde \pi   + \frac{1}{2 a^2} \alpha_\psi^2 \partial_i \tilde \pi   \partial_j \tilde \pi  \partial^2 \tilde \gamma_{ij} \right)  \\
& = \frac{2 H \tilde m_4^2 M_{\text{Pl}}^2 ( m_3^3 +4 H \tilde m_4^2)}{( m_3^3 -2 H M_{\text{Pl}}^2)^2} \int \text d^4 x \,  a \,   \dot{\tilde \pi}  \dot{\tilde \gamma}_{ij} \partial_i \partial_j \tilde \pi  
\end{align}
where as always we are in the high energy limit, and we have defined the coefficients for the constraint fields $\delta N = \alpha_{ N} \dot{\tilde \pi}$ and $\tilde \psi = \alpha_\psi \tilde \pi$ with
\begin{align}
\alpha_{ N} = \frac{ m_3^3}{ m_3^3 - 2 M_{\text{Pl}}^2 H} \ , \quad \text{and} \quad \alpha_\psi = - \frac{ m_3^3 + 4 H \tilde m_4^2}{m_3^3 - 2 H M_{\text{Pl}}^2} \ ,
\end{align}
which allows us to see more precisely where each term comes from.    The contribution from $ S_4$ comes both from the Stueckelberg discussed after \eqn{eq:term4}, and from $\delta K_{ij} \delta K^{ij}$ in the same manner as just discussed for the Einstein-Hilbert term.  More specifically, we have 
\begin{align}
 S_4 & \supset \tilde m_4^2 ( 1 - \alpha_{ N} ) (2 + \alpha_\psi)  \int \text  d^4 x \,  a \,  \dot{\tilde \pi}  \dot{\tilde \gamma}_{ij} \partial_i \partial_j \tilde \pi   = \frac{2 H \tilde m_4^2 M_{\text{Pl}}^2 ( -m_3^3 + 4 H \left[  \tilde m_4^2 +M_{\text{Pl}}^2) \right]}{(m_3^3 - 2 H M_{\text{Pl}}^2)^2}   \int \text d^4 x \,  a \,   \dot{\tilde \pi}  \dot{\tilde \gamma}_{ij} \partial_i \partial_j \tilde \pi  \ .
\end{align}
In total, then, we have 
\be \label{tildevertex}
 S_{\tilde \gamma \tilde \pi \tilde \pi} =  \frac{8 H^2 M_{\text{Pl}}^2 \tilde m_4^2 ( 2 \tilde m_4^2 + M_{\text{Pl}}^2)}{(m_3^3 - 2 H M_{\text{Pl}}^2)^2}    \int \text d^4 x \,  a \,   \dot{\tilde \pi}  \dot{\tilde \gamma}_{ij} \partial_i \partial_j \tilde \pi  \ .
\ee
Indeed, one can check that this is the same result that one would obtain by starting with the vertex in Newtonian gauge \eqn{gammapipi} and using \eqn{changeinpi} to write it in the spatially flat gauge.

\section{\label{app:gamma2pi}Graviton decay into $\gamma \pi$}

\subsection{Interaction $\gamma \gamma \pi$}

To compute the cubic vertex of the interaction $\gamma \gamma \pi$ in \eqref{starting_action}, we proceed analogously to what we did for $\gamma \pi \pi$. 
Let us start once more from the Einstein-Hilbert term, eq.~\eqref{eq:EHaction}. Focussing on the terms containing $\gamma_{ij}$, it is easy to verify that 
\be
{}^{(3)}\!R \supset -\frac{1}{4}\left(\partial_k \gamma_{ij}\right)^2 + \mathcal O(\gamma^3) \;, \qquad  E_{ij}E^{ij}  \supset  \frac{1}{4}\left(\dot\gamma_{ij}\right)^2 + \mathcal O(\gamma^3) \;,
\ee
while $E^2 \supset  \mathcal O(\gamma^3)$. The terms coming from applying the Stueckelberg trick  to these quantities would give  too many $\pi$'s to contribute to the cubic vertex. Therefore, using $N = 1-2 \Phi + {\cal O} (\Phi^2)$, the Einstein-Hilbert term contributes with
\be
S_{\rm EH}= -\frac{\MP^2}{8}\int \text d ^4 x  \; \Phi \left[\left( \partial_k \gamma_{ij}\right)^2    +\left(\dot\gamma_{ij}\right)^2 \right]  \label{eq:sEHggp} \;.
\ee
Analogously we can compute the contribution from the operator $\tilde m_4^2$, eq.~\eqref{eq:term4}. We find
\begin{align}
 S_{4} = \frac{\tilde m_4^2}{8}\int \text d ^4 x \, (2\Phi - 2\dot \pi) \left[-\left( \partial_k \gamma_{ij}\right)^2  + \left(\dot\gamma_{ij}\right)^2  \right]\label{eq:S4ggp} \;.
\end{align}

Combining these two contributions and replacing $\Phi$ by using eq.~\eqref{Phitopidot},
we obtain
\be
S_{\gamma \gamma \pi}  = -  \frac{\MP^2 \tilde m_4^2}{2(\MP^2 + 2 \tilde m_4^2)} \int \text d ^4 x  \,  \dot \pi \dot \gamma_{ij}^2 \;.
\ee
Note that, despite appearances, this vertex does not change the speed of propagation of gravitons, even in the presence of a background of $\dot \pi$. Indeed, this vertex comes from the contribution in eq.~\eqref{eq:S4ggp}, which just modifies the normalization of $\gamma$, and from the contribution of the  Einstein-Hilbert term, eq.~\eqref{eq:sEHggp}.
The latter expresses the coupling between the  kinetic terms of the graviton and the scalar metric $\Phi$, which  deforms the graviton-cone. But the same coupling and deformation are also experienced by minimally (or conformally) coupled photons and matter, so that at the end gravitons travel on the light-cone.

In terms of canonically normalized fields, the interaction vertex becomes
\begin{align}
 L _{\gamma \gamma \pi}  = - \frac{1}{ \Lambda_{\gamma\gamma\pi}^2}\dot \pi^{(c)} (\dot{\gamma}_{ij}^{(c)})^2  \label{eq:interactionggp} \;,
\end{align}
with 
\begin{equation}\label{eqLambda}
\Lambda_{\gamma\gamma\pi}^2 \equiv \frac{\MP}{\sqrt{2} \, \tilde m_4^2 } \left[ 3 m_3^6 +4 \MP^2(c + 2m_2^4)\right]^{\frac{1}{2}} \;.
\end{equation}
Denoting by $p^\mu$, $k_1^\mu$ and $k_2^\mu$ respectively the 4-momentum of the decaying graviton, of the $\pi$ field and of the graviton in the final state, in  diagrammatic form the vertex reads
\begin{align}\label{eq:vertexggp}
\vcenter{\hbox{\includegraphics[width=0.2\textwidth]{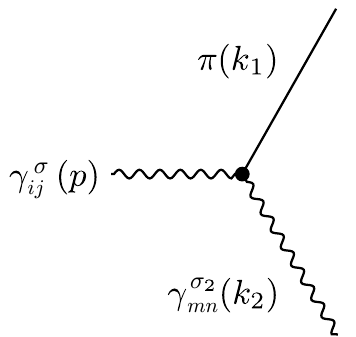}}}   = -  \frac{2 i^4}{\Lambda_{\gamma\gamma\pi}^2} E_p E_{k_1} E_{k_2}  \left[\frac{1}{2}\left(\delta_{im}\delta_{jn} + \delta_{in} \delta_{jm} \right)- \frac{1}{3}\delta_{ij}\delta_{mn}\right].
\end{align}
Note that this vertex has fewer derivatives than the vertex for $\gamma \pi \pi$, see eq.~\eqref{eqVertex2}, and the scale $\Lambda_{\gamma\gamma\pi}$ is much larger than $\Lambda_*$ defined in eq.~\eqref{Lambdadef}, i.e.~$\Lambda_{\gamma\gamma\pi} \sim \Lambda_2 \gg \Lambda_* \sim \Lambda_3$. Thus, we expect a smaller decay rate than the one from $\gamma \pi \pi$ and a weaker constraint on $\tilde m_4^2$. We will come back to this point at the end of the section.

\subsection{Decay rate}

The decay rate reads
\be
\Gamma_{\gamma \to \gamma \pi}  = \frac{1}{2 E_p } \int \frac{\text d ^3 \vect k_1}{(2 \pi)^3 2 E_{k_1}}\frac{\text d ^3 \vect  k_2}{(2 \pi)^3 2 E_{k_2}} (2 \pi)^4 \delta^{(4)}(p^\mu - k_1^\mu - k_2^\mu) \braket{|i \Ccc A|^2} \;,
\ee
where $\braket{|i \Ccc A|^2}$ is the the matrix element squared and averaged over the polarizations of the initial and final states.

As done in Sec.~\ref{subsec:decayrate}, we can 
remove   $\delta^{(3)} (\vect p - \vect  k_1 - \vect k_2)$ by integrating over $\text d^3 \vect k_2$. Moreover, integrating over $\text d k_1$ 
using the on-shell conditions
\be
E_p = p \;, \qquad E_{k_1} = c_s k_1 \;, \qquad E_{k_2} = k_2 \;,
\ee
we can remove $\delta(E_p- E_{k_1} - E_{k_2})$. To do that, we express $k_2$ in terms of $k_1$ and $\Omega = \vect p \cdot \vect k_1 /(p k_1)$ using eq.~\eqref{k2fork1}. 
In the following we assume $0<c_s \le \Omega$; the case $c_s > \Omega$, and thus $c_s >1$, is kinematically forbidden. Replacing $k_1$ using
\begin{align}
k_1 = \frac{2 p (\Omega-c_s)}{1-c_s^2} \label{Solk1}\;,  
\end{align}
we obtain
\be \label{eqGammaggpimp}
\Gamma_{\gamma \to \gamma \pi}  = \frac{1}{2 p } \frac{1}{4 \pi c_s (1-c_s^2)} \int_{c_s}^{1} \text d \Omega \, \braket{|i \Ccc A|^2}  \;.
\ee

Let us now compute $\braket{|i \Ccc A|^2}$. This is given by 
\begin{align}\label{eqAmp}
\braket{|i \Ccc A|^2} = \frac{1}{2}\sum_{\sigma = \pm} \sum_{\sigma_2 = \pm} |i \Ccc A|^2 \;,
\end{align}
where the tree-level amplitude reads
\begin{equation}
i \Ccc A =- \frac{2}{\Lambda_{\gamma\gamma\pi}^2} E_p  E_{k_1} E_{ k_2}  \, \epsilon_{ij}^{\star \, \sigma}(\vect p) \epsilon_{ij}^{ \sigma_2}(\vect k_2) \;.
\end{equation}
Using this expression, eq.~\eqref{prodeps} and the transversality condition,
after some straightforward algebra we find
\begin{align}\label{eqAmp2}
\braket{|i \Ccc A|^2} =  \frac{2}{\Lambda_{\gamma\gamma\pi} ^4} (c_s \, p \, k_1  k_2 )^2  \left[ 3 +  6 \frac{(\vect{k}_2 \cdot \vect{p})^2}{\vect{k}_2^2 \vect{p}^2} + \frac{(\vect{k}_2 \cdot \vect{p})^4}{\vect{k}_2^4 \vect p^4} \right] \;.
\end{align}

After we replace this result in \eqref{eqGammaggpimp} and perform the integral over $\text d \Omega$ we find
\begin{equation}
\Gamma_{\gamma \to \gamma \pi}  = \frac{p^5}{32 \pi \Lambda_{\gamma\gamma\pi}^4}\mathcal F(c_s) \;,
\end{equation}
where 
\begin{align}
\mathcal F(c_s) &\equiv -\frac{1}{c_s^{10}}(1-c_s^2)^3 (5 - c_s^2) (\text{tanh}^{-1} \, c_s +1) +\frac{1}{c_s^{10}(1+c_s)^5}\left[5+30 c_s+59 c_s^2+\frac{17 c_s^3}{3}-\frac{416 c_s^4}{3}  -\frac{509
   c_s^5}{3}\right. \nonumber \\
& \left.+22 c_s^6+\frac{6458 c_s^7}{35}+\frac{2329
   c_s^8}{21}-\frac{10496 c_s^9}{315}-\frac{3791 c_s^{10}}{63}+\frac{3743
   c_s^{11}}{105}-\frac{3782 c_s^{12}}{63}+\frac{3263 c_s^{13}}{63}\right]\; .
 \end{align}
 The function $\mathcal F(c_s)$ vanishes for $c_s = 0$ and reaches its maximum value $\mathcal F(c_s^{\text{max}}) \approx 3.50$ at $c_s^{\max} \approx 0.19$. Additionally, $\mathcal F(1) = 4/3$.

Applying this result to LIGO/Virgo energies, $p \sim \Lambda_3  $, and requiring that the decay  is slower than a Hubble time one gets
\be
\left( \frac{\Lambda_3}{\Lambda_2} \right)^2 \left( \frac{\Lambda_2}{{\Lambda_{\gamma\gamma\pi}}} \right)^4 \frac{\mathcal F(c_s)}{ 32  \pi } \sim 10^{-20} \left( \frac{\Lambda_2}{{\Lambda_{\gamma\gamma\pi}}} \right)^4  \lesssim 1\;.
\ee
Since 
\be
\left( \frac{ \Lambda_2}{\Lambda_{\gamma\gamma\pi}} \right)^4 = \,\frac{\alpha_{\rm{H}}^2}{4 \, \alpha}\;,
\ee
the constraint on $\alpha_{\rm H}$, and thus on $\tilde m_4^2$, is rather weak.



\small{
\bibliography{bib_v3} }
\bibliographystyle{utphys}

\end{document}